\documentclass{article}
\usepackage[utf8]{inputenc}
\usepackage{caption}
\usepackage{graphicx}
\usepackage{float} 
\usepackage{subcaption}
\usepackage{cite}
\usepackage[namelimits]{amsmath} 
\usepackage{amssymb}    
\usepackage{amsmath}
\usepackage{enumerate}
\usepackage[colorlinks,linkcolor=blue]{hyperref}
\usepackage{ulem}

\title{Collective action and spontaneity cycles: Cascading dynamics under Bayesian games}

\author{FangYiKuang,Ding¹{ }{ }Jie,Xu²{ }{ }Mingke,Li³\\
¹²³ Department of sociology\\
East China University of science and technology\\ Shanghai\\
China}
\date{June 2022}

\begin{document}

\maketitle

\begin{abstract}
The formation mechanisms and cyclical conditions of collective action have become open issues in research involving public choice, social movements, and more. For this reason, on the basis of rational decision-making and social assimilation, this paper proposes an action model that combines Bayesian game and social network dynamics, and incorporates exogenous cycles into it. For this model, this paper proves the spontaneous action theorem and action cycle theorem of collective action, and based on numerical simulation and empirical calibration, further confirms the theoretical mechanism involving elements such as risk/risk-free incentives and the number of social ties. Based on such conclusions and evidence, this paper proposes a theory of spontaneous cycles as an integrative answer to the open question of collective action formation/cycles. \\

\textbf{keywords:}\quad collective action, cycle, social assimilation, threshold, spontaneity
\end{abstract}

\section{Introduction}
\subsection{Questions raised}
In the fields of sociology and economics, collective action, as a concept covering important group phenomena such as political elections, social movements, and online group buying, has received extensive attention in formal theory and empirical research. Since the 1970s, due to the maturity of the democratic process and civil society, most scholars have gradually abandoned the "group behavior theory" and regarded related events as "emotion/violence/blind obedience" and explored collective action in different approaches, such as the creation of public goods (Olson and public choice theory), the combination of market resource mobilization and political process (Taro et al.) Regular social movements (racial equality/environment/lgbt), team decision-making and innovation, and establish institutional middle-level theories.\\

However, in the above-mentioned fields, a kind of spontaneous but limited institutionalized collective action has emerged in developed and southern countries (later it will be strictly referred to as spontaneous collective action), and it has a rich macro background. It is the "labor resistance movement"\cite{1}\cite{23}, which includes various forms such as strikes and demonstrations, and occupies an important position in cluster events such as Occupy Wall Street and the Yellow Vest Movement. As far as China is concerned, it involves the outbreak of multiple incidents such as construction workers demanding wages, truck drivers/delivery workers strike, Zhengzhou Foxconn demonstration, etc., and is embedded in the operation of civil politics and urban-rural dual-track economy. At the same time, in this type of collective action sequence, we can also observe a complex fluctuation—the number of protests in a specific industry is in a cyclical ups and downs, reflecting the existence of “periodicity”. This attribute also appears in various social movements such as Memorial Day parades and feminist movements. Based on this, we can regard labor protests as an important field of vision for understanding collective action in the following texts, and raise two basic questions in this paper, exploring the causal law and mechanism of a single outbreak of spontaneous collective action and periodic fluctuations, which are presented as follows:
\begin{enumerate}[(1)]
\item What are the formation conditions and causal mechanisms of collective action (question 1/short-term and mesoscopic phenomena)?
\item How do collective action cycles arise (Question 2/Long-term macro-phenomena)?
\end{enumerate}

\subsection{Literature review}
\subsubsection{Collective action: the case of the labor movement}
In order to understand a type of collective action with spontaneity, we take the local labor movement as an example and review the existing literature, which is often expressed as "labour mass events". Regarding the causes and mechanisms of such events/actions, relevant research approaches can be divided into two categories: one focuses on macro factors such as class, labor market, labor law/arbitration system, and national governance; the other focuses on the impact of micro-attributes such as cognitive biases and motivating emotions (such as relative deprivation) on action participation from the perspective of individuals. Therefore, based on these two perspectives, we can summarize and explain the literature involved.\\

As far as macro elements are concerned, we can discuss them separately from social stratification, labor supervision system and state-society relationship. Among them, the uneven resource input/ambiguity of property rights involved in class differentiation is regarded as the cause of conflict, while the obstruction of class mobility and insufficient public relief further promote labor participation actions\cite{10}\cite{11}\cite{12}; in terms of labor supervision, the failure of trade unions and the lack of arbitration system can prevent workers’ interest demands from being handled within the institutional framework, prompting them to turn to social movement models\cite{2}; in state-society relations On the one hand, the characteristics of labor protests involve the logic of maintaining stability, rigid governance models, and the weakness of existing social forces (such as the lack of coordinated NGOs)\cite{5}\cite{12}, the latter making it difficult to institutionalize, and retaining some characteristics similar to spontaneity and insufficient organization.\\

As far as the micro elements are concerned, cognition-emotion, income analysis and relationship network may be three core perspectives. On the cognitive-emotional level, individuals are motivated to participate in actions\cite{1} due to their cognitive biases towards things such as event responsibility attribution/behavioral outcome expectations\cite{1}. On the basis of individual rationality and income analysis, the core factor of individual action participation lies in the "difficulty of expressing interests", such as the lack of enterprise negotiation mechanism\cite{4}, and embedded in the reformed labor-capital structure, so that workers have "common benefits". Therefore, individuals can protect their own interests by establishing cooperative labor movements (such as township strikes)\cite{9}. In terms of relational networks, the diffusion of labor group incidents can be effectively explained, such as emphasizing the relationship connection role of mass media, the coupling effect of virtual/communication/face-to-face networks, etc.\cite{6}\cite{7}, which can explain the mobilization methods of labor movements.\\

However, in general, the above-mentioned attributions to labor movements are often based on partial measurement/experimental results or field observations, failing to provide the “micro-macro-integrated explanation” needed in this paper, and to bridge the gap with general collective action theories. In this regard, we may need to pay attention to two related processes: one is the discussion of rational decision-making (such as incomplete information games), which involves interest demands, information asymmetry and public choice issues involved in individual participation in actions such as strikes (such as joint benefits/joint participation can achieve goals), which can integrate the above content about conflicts of interest/institutional characteristics; the second is the study of social assimilation. Complex social bonding, perspective learning, and imitative behavior, which can incorporate the above discussion of social cognitive biases/deprivation/relational connections. Both of these two processes can be brought into the perspective of "collective action" for further excavation and discussion.

\subsubsection{Formation of collective action}
Based on the above analysis, we can briefly review the theoretical perspectives in collective action and provide a basis for establishing a formal model below. In this field, the mechanism of individual participation in action can include two typical levels: (1) At the level of single individual decision-making, the incentives received by individuals are closely related to their utility (for example, whether they participate in a strike or not will successfully ask for wages), so it is an important agenda to conduct research on different types of incentives; (2) At the level of interaction between individuals, there are two directions—one is the formation and dissemination of cognitive frameworks. Taking the labor movement as an example, this includes the diffusion of ideas such as equal wages and labor safety and security, which can be called dynamic “framing”; The second is the opportunities, resources, and organizational structure that collective action relies on. In this case, it involves labor law and trade union organizations, specific industry markets and wage conditions, and the synergy between labor NGOs and industry associations. It can be regarded as a combination of "resource mobilization"/"political process"\cite{13}. On the whole, these two are related to the "rational decision-making" and "social assimilation" mentioned above respectively, and they try to study the relevant conditions, environments and mechanisms: we will explain them below.\\

As far as rational decision-making is concerned, Olson’s classic work laid the direction for follow-up research: in the analysis of collective action, he found two types of sources related to individual utility, namely “collective incentives for public goods” and “selective incentives”\cite{14}—the former represents the resources that both participants and non-participants can enjoy after successful actions, while the latter belongs to the benefits bestowed on specific types of individuals (such as active participants). In the fields of collective market action/world economics, this division has been embodied and expanded in multiple ways, including "different cultural elements in products have different levels of incentives for individuals with different ideologies"/"introducing utility coefficients in the global game/distinguishing the different utility levels of non-participants and participants after successful actions"\cite{15}\cite{16}\cite{17}. However, another kind of utility division is worthy of attention: for individuals, some utility can only be obtained by “self-participation in action” (such as wages obtained by asking for wages/compensation for “distributed according to trouble”), while some utility requires “simultaneous participation of others” (such as improving collective working hours and working environment, etc.). The former/the latter can be called “private utility”/“common utility” respectively. In this regard, the following will refer to this classification to analyze the two types of incentives/utilities received by individuals and their effects.\\

As far as social assimilation is concerned, at the level of framework construction, we can divide it into two levels according to the cases of the labor movement\cite{13}: one is the formation and diffusion of cognitive frameworks such as labor rights protection/social movement mutual assistance, and the other is the competition between it and different cognitive frameworks—the former often includes "diagnosis" and "motivation", which deal with the tasks of "constructing labor issues" (such as construction site/factory safety) and "proposing an action agenda" (such as increasing labor protection measures/increasing compensation for accidents) respectively The latter involves the issue of "anti-framing" and negative attitudes towards the labor movement, such as questioning "civil disobedience undermines social stability"/"mass incidents are distributed according to disturbances and undermines fairness". The latter will be reflected in the models that follow, as perceived resistance/"thresholds to action"\cite{20} felt by actors, fused with cascading dynamics about collective action\cite{21} At the level of resource mobilization and political process, it is important to understand collective action as a market process and understand its political opportunities\cite{15}. For the former, it may be combined with Olson’s insights: for collective action participation, individuals can be subject to two types of incentives (private utility/common utility), and choose whether to take participatory actions under specific rules based on the inference of each other’s characteristics (private information). For the latter, policy incentives for collective action constitute a typical source of opportunity. Such elements are modeled below as a Bayesian game\cite{16} to incorporate both the characteristics of incomplete information and the effects of political opportunity in social movements.\\

Based on the discussion of collective action above, we can decompose Question 1 into two: How is collective action affected by the two types of utility/incentives (denoted as Question 1-1/emphasis on rationality)? How does its facilitation relate to cognitive interactions and social networks between individuals (denoted as Question 1-2/emphasis on assimilation)? Clearly, Questions 1-1/1-2 attempt to integrate Olson's framework with insights from the social network school and apply this to the answer to spontaneous collective action. In addition, for questions 1-1/1-2, we can also supplement and examine the existing conclusions—as far as 1-1 is concerned, in addition to the selective incentives/public incentives that Olson divides, there are other core types of incentives that affect collective action (that is, incentives corresponding to private/common utility, called risk-free incentives/risk incentives), which will be discussed in this article; as far as 1-2 is concerned, the existing concepts of the social network school need to be challenged, that is, "high social ties lead to high social capital, and high social capital leads to action participation."\cite{16}, this concept is refuted and reconstructed in the following proof/simulation and theoretical construction.

\subsubsection{Collective action cycle}
After a basic literature review on collective action formation, we can transition to Question 2, summarizing the relevant research on action cycles. As far as the definition of this concept is concerned, it includes a wide variety of terms, including "protest cycle" (protest cycle), social movement cycle (social movement cycle), etc., mainly referring to "a phenomenon that occurs cyclically in a specific collective action process" (such as arousal-recession, etc.). In this regard, the existing literature mainly involves cyclical phenomena in specific regions, such as European revolutions and labor movement waves, etc., and mainly proposes empirical conclusions rather than middle-level theoretical frameworks. They can be divided into "regular explanations" and "mechanistic explanations" for action cycles, which are discussed one by one below.\\

In terms of laws, the factors that affect the cycle of collective action can be divided into "variables within the system" and "variables outside the system". The former describes the attributes of action participants and the environment in which they live. As far as protests and social movements are concerned, this may include the cyclical characteristics of the policy-making process itself (bargaining with social movement agendas), as well as fluctuations in "political opportunities" such as public opinion attention\cite{24}\cite{13}. The latter explores the relationship between action cycles and other system cycles (such as economic cycle/anniversary cycle/organizational life cycle, etc.), which often involves the evolution and fluctuation of individual occupations, classes, and life opportunities over time\cite{25}in the classic perspective of Tarot and others.\\

As far as the mechanism is concerned, we can discuss the two perspectives of "collective decomposition" and "stage decomposition"\cite{26}respectively. The former involves an analysis of the roles of participants in collective action, which may include protesters, oppressors, and communicators, and explains the "ebb and flow" of actions through multiple interactions among the three types of subjects (for example, communicators can both trigger struggles and resolve conflicts), and focus on the "diffusion process" of actions participating in the crowd\cite{26}. The latter attempts to divide an action cycle into multiple stages. The most intuitive classification criteria for this are the “rise” (period) and “decline (period)” of collective action: some studies have found that material abundance and post-materialist values can lead to the rise of collective action over a long period of time; the mixture of radicalization and organization can lead to the decline of collective action; this is due to “high activity of the leisure class” and “insufficient action power caused by ideological differences”\cite{24}. Generally speaking, although these two methods lack generally reliable mechanism verification and comprehensive explanation, they can provide method reference for the analysis of the cycle in the following sections.\\

\section{Models and Theorems}
\subsection{Overview}
Based on the literature review in the previous part, we can integrate rational decision-making/social assimilation/cycle elements to establish a formal model describing collective actions such as strikes. Since this model is not only applicable to specific cases (such as participation in workers' protests), but also describes a general type of general collective action (i.e. spontaneous collective action which will be strictly defined later), we will use the terms in these two objects at the same time below to give a semantic interpretation of the model. As these terms can basically be transformed into each other (specific/general), they should not cause misunderstandings, such as "collective action participation-participation in workers' protest", "activists-labor activists", "probability of willingness to act-probability of willingness to protest" and so on. The core content of the model can be summarized in three parts:\\

(1) At the level of rational decision-making, we try to obtain the long-term equilibrium probability of i choosing to act or not choosing to act according to the type of individual i (whether active), belief (whether others are active) and utility (income from participating in actions)—here only considers the labor movement in a specific industry. Therefore, we can establish a Bayesian game $(I, \theta, P, S, U)$ and use its mixed strategy Nash equilibrium (the long-term equilibrium probability of choosing to act or not to act) as an answer to this question, which is in line with the previous analysis of "rational decision-making" (such as utility can reflect the interests of workers).\\

(2) At the level of social assimilation, we can introduce an averaging dynamics\cite{18} of opinion evolution-collective action fusion, and incorporate the long-term equilibrium probability described in (1). Its overall idea is that at time t, each individual i simultaneously learns the negative evaluation of people around him about participating in the struggle (threshold T/action resistance) and the action ratio of people around him (proportion perception P/action thrust), and performs a weighted comparison between P and T according to his own equilibrium probability. When the thrust is greater than the resistance, i will participate in collective action, otherwise, it will not participate in it, which is derived from the previous discussion of "social assimilation" (such as the threshold represents the frame/anti-frame, etc.).\\

(3) At the periodic level, we introduce an exogenous willingness coefficient f(t) as an exogenous variable that affects the individual's willingness to act. This function can have periodicity/pan-periodicity (such as wavelets). In the case of the labor movement, it can reflect the impact of economic conditions and fixed contract time limits (such as signing at the beginning of the year/paying at the end of the year) on the internal system of collective action. Because it incorporates factors such as the economy and has a fluctuating attribute, this can echo Tarrow's research on cycles mentioned above, and it is related to the perspective of the mechanism of action diffusion in the follow-up analysis.\\

Aiming at this compound model of game-dynamics, we can directly cut into the research question of this paper, that is, "the formation conditions/mechanisms and periodic causes of collective action". For the former (the emergence of large-scale collective action), we can translate collective action into a term in dynamical systems, called "synchronization" - its semantic interpretation is that the state of each individual in the system tend to be consistent. Thus, if there are individuals acting and the system is synchronized, then the phenomenon induced by the system is collective action (every individual acts). Based on this connection, we explore below the sufficient conditions for models to achieve synchronization (collective action), which can be expressed as parameter ranges for properties such as private incentives/social networks. For the latter, the collective action cycle is equivalent to the "synchronous cycle" in the model, that is, the cycle in which the event "all/most people participate in action" occurs, which will also be paid attention to in the following theorems/simulations.

\subsection{Formal model}
\subsubsection{Game framework}
According to the elaboration on the level of rational decision-making above, we can consider a five-tuple $\left( I,\Theta ,P,S,U \right)$ representing a static Bayesian game, which means:

\begin{description}
\item[{\large\textcircled{\footnotesize 1}}] I=$\{i=1,j=2\}$ represent a set of two potential collective action participants.
\item[{\large\textcircled{\footnotesize 2}}] ${{\Theta }_{\text{i}}}={{\Theta }_{\text{j}}}=\left\{ {{\theta }_{1}}=1,{{\theta }_{2}}=-1 \right\}$is the type set of individual i and individual j, 1 indicates that the individual is active (such as a labor activist), and -1 indicates that the individual is not active(such as non-labor activist).
\item[{\large\textcircled{\footnotesize 3}}] ${{P}_{\text{i}}}\left( {{\theta }_{\text{1}}}\left| {{\theta }_{\text{k}}} \right. \right)=x\left( \forall {{\theta }_{\text{k}}}\in {{\Theta }_{1}} \right);{{P}_{\text{j}}}\left( {{\theta }_{1}}\left| {{\theta }_{\text{k}}} \right. \right)=y\left( \forall {{\theta }_{\text{k}}}\in {{\Theta }_{2}} \right)$means that the probability that individual i/j thinks the other j/i is active is x and y respectively, and can be set.
\item[{\large\textcircled{\footnotesize 4}}] ${{S}_{\text{i=1}}}={{S}_{\text{j=1}}}=\{B=1({{S}_{1}}),NB=-1({{S}_{2}})\}$represent the strategy set (the same) of individual i and individual j, B and NB represent the two strategies of participation(acting)/non-participation(not acting) in collective action respectively.
\item[{\large\textcircled{\footnotesize 5}}] ${{u}_{i=1}}\left( {{\theta }_{i}},{{\theta }_{j}},{{\text{S}}_{i}},{{S}_{j}} \right)$\\
$=\alpha \cdot \text{I}{{\text{I}}_{\left\{ 1 \right\}}}\left( {{S}_{i}} \right)\left( 1-\left| {{\theta }_{i}}-{{S}_{i}} \right| \right)+\beta \cdot \text{I}{{\text{I}}_{\left\{ 1 \right\}}}\left( {{S}_{i}} \right)\cdot \text{I}{{\text{I}}_{\left\{ 1 \right\}}}\left( {{S}_{j}} \right)$
${{u}_{j=2}}\left( {{\theta }_{i}},{{\theta }_{j}},{{S}_{i}},{{S}_{j}} \right)$\\
$=\alpha \cdot \text{I}{{\text{I}}_{\left\{ 1 \right\}}}\left( {{S}_{j}} \right)\left( 1-\left| {{\theta }_{j}}-{{S}_{j}} \right| \right)+\beta \cdot \text{I}{{\text{I}}_{\left\{ 1 \right\}}}\left( {{S}_{i}} \right)\cdot \text{I}{{\text{I}}_{\left\{ 1 \right\}}}\left( {{S}_{j}} \right)$
\end{description}
The above are the utility functions of individual i / individual j. The two are symmetrical to each other, and the specific meanings can be discussed one by one, namely (1) For individual i(or j), if he doesn't participate in collective action(not act), the utility is 0; (2) If he participates in the action(act), there are four possibilities: A. I am an active actor and the other party does not participate in the action, the utility is $\alpha$, B. I am a non-active actor and the other party does not participate in the action, the utility is $-\alpha$, C. If you are an active actor and the other party also participates in the action, the utility is $\alpha + \beta $, D. If you are a non-active actor and the other party also acts, the utility is $-\alpha+\beta$. Among them, $\alpha$ represent the degree of private utility (such as the value of individual salary/"promoting the concept of workers' rights/participation in protest" to labor activists), $\beta$ represent the magnitude of common utility (such as the improvement of the labor environment after the success of collective action, etc.).
\\
Based on the above model, we can induce a Bayesian mixed-strategy Nash equilibrium\cite{22} is the mixed strategy Nash equilibrium solution of the "Selten game" corresponding to the Bayesian game], which stands for "when individual i is type k, he/she chooses S1/S2 with this probability". Specifically, for $\left( i,k \right)$ of individual $i$ type $k$, there is the probability $\left( {{P}_{ik,1}},{{P}_{ik,2}} \right)$ of ${{S}_{1}}/{{S}_{2}}$, known from the symmetry of the game, it has nothing to do with, only related to the type $k$, so the i-th type Type $i$ can be selected as s1 , the probability of s2 (probability of willing to act and willing not to act) is recorded as $\left( {{P}_{T{{\text{y}}_{\text{i}}},1}},{{P}_{T{{\text{y}}_{\text{i}}},2}} \right)$.

\subsubsection{Collective action dynamics}

\paragraph{Environment and initial value setting}

\paragraph{Update rules and synchronization}

\subsection{Main theorems}
In this section, we present theoretical results obtained from the above model: where Theorem 1/Theorem 2 answer Question 1 (collective action formation) and Question 2 (collective action cycle), respectively. Considering that the rigorous formulation and proof of theorems are cumbersome, we have added “remarks” below Lemma 1/Lemma 2/Theorem 1/Theorem 2 respectively to illustrate the sociological implications of these propositions.\\

\textbf{Lemma1:}Under M1, the mixed Nash equilibrium induced by $\left( I,\theta ,P,S,U \right)$ is $\left( {{P}_{1}},{{P}_{2}},{{P}_{3}},{{P}_{4}} \right)$, where ${{P}_{1}}={{P}_{3}}=(a/b)[(x+y-2)/(x-y)]$, ${{P}_{2}}={{P}_{4}}=(a/b)[(x+y)/(x-y)]$.\\

\textbf{prove:}
\\(1) According to Selten, the game can be transformed into a strategy type $\left( {I}',{S}',{U}' \right)$, respectively\\

{\large\textcircled{\footnotesize 1}} ${I}'=I\times \theta =\left\{ \left( i{{\theta }_{1}} \right),\left( i,{{\theta }_{2}} \right),\left( j,{{\theta }_{1}} \right),\left( j,{{\theta }_{2}} \right) \right\}$, and $\left( i,{{\theta }_{k}} \right)/(j,\left. {{\theta }_{k}} \right)$ are hereinafter denoted as ${{i}_{k}}/{{j}_{k}}$\\

{\large\textcircled{\footnotesize 2}} ${S}'=S\times S=\prod\limits_{i=1}^{4}{\left\{ {{s}_{1}},{{s}_{2}} \right\}}$\\

{\large\textcircled{\footnotesize 3}} $U$ is expressed as: ${{S}_{m}},{{S}_{{{\text{n}}_{\text{1}}}}},{{S}_{{{\text{n}}_{\text{2}}}}}$ any $k=1,2$ yes, $k=1$, \\

${{u}_{{{i}_{k}}}}\left( {{S}_{m}},{{S}^{*}},{{S}_{{{n}_{1}}}}\cdot {{S}_{{{n}_{2}}}} \right)$\\

$={{P}_{1}}\left( {{\theta }_{1}}\mid {{\theta }_{k}} \right)\cdot {{U}_{i}}\left( {{\theta }_{k}},{{\theta }_{1}},{{S}_{m}},{{S}_{{{n}_{1}}}} \right)+{{P}_{1}}\left( {{\theta }_{2}}\mid {{\theta }_{k}} \right)\cdot {{U}_{i}}\left( {{\theta }_{k}},{{\theta }_{2}},{{S}_{m}},{{S}_{{{n}_{2}}}} \right)$\\

$=x\cdot \left[ \alpha \cdot \text{I}{{\text{I}}_{\left\{ 1 \right\}}}\left( {{S}_{m}} \right)\left( 1-\left| {{\theta }_{k}}-{{S}_{m}} \right| \right)+\beta \cdot \text{I}{{\text{I}}_{\left\{ 1 \right\}}}\left( {{S}_{m}} \right)\cdot \text{I}{{\text{I}}_{\left\{ 1 \right\}}}\left( {{S}_{{{n}_{1}}}} \right) \right]+(1-x)[\alpha \cdot $\\

$\left. \text{I}{{\text{I}}_{\left\{ 1 \right\}}}\left( {{S}_{m}} \right)\left( 1-\left| {{\theta }_{k}}-{{S}_{m}} \right| \right)+\beta \cdot \text{I}{{\text{I}}_{\left\{ 1 \right\}}}\left( {{S}_{m}} \right)\cdot \text{I}{{\text{I}}_{\left\{ 1 \right\}}}\left( {{S}_{{{n}_{2}}}} \right) \right]$\\

$=\alpha \cdot \text{I}{{\text{I}}_{\left\{ 1 \right\}}}\left( {{S}_{m}} \right)\left( 1-\left| {{\theta }_{k}}-{{S}_{m}} \right| \right)+\beta \cdot \text{I}{{\text{I}}_{\left\{ 1 \right\}}}\left( {{S}_{m}} \right)\cdot \left[ x\cdot \text{I}{{\text{I}}_{\left\{ 1 \right\}}}\left( {{S}_{{{n}_{1}}}} \right)+(1\cdot x)\cdot \text{I}{{\text{I}}_{\left\{ 1 \right\}}}\left( {{S}_{{{n}_{2}}}} \right) \right]$\\

${{u}_{{{j}_{k}}}}\left( {{S}_{{{m}_{1}}}},{{S}_{{{m}_{2}}}},{{S}_{n}},{{S}^{\infty }} \right)$\\

$={{P}_{2}}\left( {{\theta }_{1}}\mid {{\theta }_{k}} \right)\cdot {{u}_{i}}\left( {{\theta }_{1}},{{\theta }_{k}},{{S}_{{{m}_{1}}}},{{S}_{n}} \right)+{{P}_{2}}\left( {{\theta }_{2}}\mid {{\theta }_{k}} \right)\cdot {{u}_{i}}\left( {{\theta }_{2}},{{\theta }_{k}},{{S}_{{{m}_{2}}}},{{S}_{n}} \right)$\\

$=x\cdot \left[ \alpha \cdot \text{I}{{\text{I}}_{\left\{ 1 \right\}}}\left( {{S}_{n}} \right)\left( 1-\left| {{\theta }_{k}}-{{S}_{n}} \right| \right)+\beta \cdot \text{I}{{\text{I}}_{\left\{ 1 \right\}}}\left( {{S}_{{{m}_{1}}}} \right)\cdot \text{I}{{\text{I}}_{\left\{ 1 \right\}}}\left( {{S}_{n}} \right) \right]+(1-x)$\\

$\left[ \alpha \cdot I{{I}_{\left\{ 1 \right\}}}\left( {{S}_{n}} \right)\left( 1-\left| {{\theta }_{k}}-{{S}_{n}} \right| \right)+\beta \cdot I{{I}_{\left\{ 1 \right\}}}\left( {{S}_{{{m}_{2}}}} \right)\cdot \text{I}{{\text{I}}_{\left\{ 1 \right\}}}\left( {{S}_{n}} \right) \right]$\\

$=\alpha \cdot I{{I}_{\left\{ 1 \right\}}}\left( {{S}_{n}} \right)\left( 1-\left| {{\theta }_{k}}-{{S}_{n}} \right| \right)+\beta \cdot I{{I}_{\left\{ 1 \right\}}}\left( {{S}_{n}} \right)\cdot \left[ x\cdot I{{I}_{1}}\left( {{S}_{{{m}_{1}}}} \right)+(1-x)I{{I}_{1}}\left( {{S}_{{{m}_{2}}}} \right) \right]$\\

$k=2$ is the same, so a strategic game can be defined.\\

(2)For $\left( {I}',{S}',{U}' \right)$, in order to find its mixed Nash equilibrium, we record the mixed strategy of ${{i}_{1}},{{i}_{2}},{{j}_{1}},{{j}_{2}}$ as $\left( {{p}_{i}},1-{{p}_{i}} \right),i=1,2,3,4$. Accordingly, taking agent ${{i}_{1}}$ as an example, the following steps can be taken:\\

{\large\textcircled{\footnotesize 1}}The expected utility of ${{i}_{1}}$ choosing ${{S}_{1}}$ is\\

${{\operatorname{Eu}}_{{{i}_{1}}}}\left( {{S}_{1}},\cdot ,\cdot ,\cdot  \right)$\\

$={{P}_{3}}\left[ {{P}_{4}}u\left( {{S}_{1}},{{S}^{*}},{{S}_{1}},{{S}_{1}} \right)+\left( 1-{{P}_{4}} \right)u\left( {{S}_{1}},{{S}^{*}},{{S}_{1}},{{S}_{2}} \right) \right]+\left( 1-{{P}_{3}} \right)\left[ {{P}_{4}}u\left( {{S}_{1}},{{S}^{*}}, \right. \right.$\\

$\left. \left. {{S}_{2}},{{S}_{1}} \right)+\left( 1-{{P}_{4}} \right)u\left( {{S}_{1}},{{S}^{*}},{{S}_{2}},{{S}_{2}} \right) \right]$\\

$={{P}_{3}}\left[ {{P}_{4}}u\left( {{S}_{2}},{{S}^{*}},{{S}_{1}},{{S}_{1}} \right)+\left( 1-{{P}_{4}} \right)u\left( {{S}_{2}},{{S}^{*}},{{S}_{1}},{{S}_{2}} \right) \right]+\left( 1-{{P}_{3}} \right)\left[ {{P}_{4}}u\left( {{S}_{2}}, \right. \right.$\\

$\left. \left. {{S}^{*}},{{S}_{2}},{{S}_{1}} \right)+\left( 1-{{P}_{4}} \right)u\left( {{S}_{2}},{{S}^{*}},{{S}_{2}}{{S}_{2}} \right) \right]$\\
$=E{{u}_{{{i}_{1}}}}\left( {{S}_{2}},\cdot ,\cdot ,\cdot  \right)$\\

Substituting into the specific utility function,\\

${{P}_{3}}\left[ {{P}_{4}}\{\alpha \cdot \text{I}{{\text{I}}_{\left\{ 1 \right\}}}\left( {{S}_{1}} \right)\left( 1-\left| {{\theta }_{k}}-{{S}_{1}} \right| \right)+\beta \cdot \text{I}{{\text{I}}_{\left\{ 1 \right\}}}\left( {{S}_{1}} \right)\cdot \left[ \text{I}{{\text{I}}_{\left\{ 1 \right\}}}\left( {{S}_{1}} \right) \right]\}+\left( 1-{{P}_{4}} \right)\{\alpha \cdot  \right.$\\

$\left. \text{I}{{\text{I}}_{\left\{ 1 \right\}}}\left( {{S}_{1}} \right)\left( 1-\left| {{\theta }_{k}}-{{S}_{1}} \right| \right)+\beta \cdot \text{I}{{\text{I}}_{\left\{ 1 \right\}}}\left( {{S}_{1}} \right)\cdot [x\cdot \text{I}{{\text{I}}_{\left\{ 1 \right\}}}\left( {{S}_{1}} \right)+(1-x)\cdot \text{I}{{\text{I}}_{\left\{ 1 \right\}}}\left( {{S}_{2}} \right)]\} \right]+$\\

$\left( 1-{{P}_{3}} \right)\left[ {{P}_{4}}\left\{ \alpha \cdot \text{I}{{\text{I}}_{\left\{ 1 \right\}}}\left( {{S}_{1}} \right)\left( 1-\left| {{\theta }_{k}}-{{S}_{1}} \right| \right)+\beta \cdot \text{I}{{\text{I}}_{\left\{ 1 \right\}}}\left( {{S}_{1}} \right)\cdot [x\cdot \text{I}{{\text{I}}_{\left\{ 1 \right\}}}\left( {{S}_{2}} \right)+(1\cdot x)\cdot  \right. \right.$\\

$\left. \left. \text{I}{{\text{I}}_{\left\{ 1 \right\}}}\left( {{S}_{1}} \right) \right] \right\}+\left( 1-{{P}_{4}} \right)\{\alpha \cdot \text{I}{{\text{I}}_{\left\{ 1 \right\}}}\left( {{S}_{1}} \right)\left( 1-\left| {{\theta }_{k}}-{{S}_{1}} \right| \right)+\beta \cdot \text{I}{{\text{I}}_{\left\{ 1 \right\}}}\left( {{S}_{1}} \right)\cdot [x\cdot \text{I}{{\text{I}}_{\left\{ 1 \right\}}}\left( {{S}_{2}} \right)$\\

$\left. \left. \left.+(1-x)\cdot \text{I}{{\text{I}}_{\left\{ 1 \right\}}}\left( {{S}_{2}} \right) \right] \right\} \right]$\\

$={{P}_{3}}\left[ {{P}_{4}}\left\{ \alpha \cdot \left( 1-\left| {{\theta }_{k}}-{{S}_{1}} \right| \right)+\beta  \right\}+\left( 1-{{P}_{4}} \right)\left\{ \alpha \cdot \left( 1-\left| {{\theta }_{k}}-{{S}_{1}} \right| \right)+\beta \cdot x \right\} \right]+$\\

$\left( 1-{{P}_{3}} \right)\left[ {{P}_{4}}\left\{ \alpha \cdot \left( 1-\left| {{\theta }_{k}}-{{S}_{1}} \right| \right)+\beta \cdot (1-x) \right\}+\left( 1-{{P}_{4}} \right)\left\{ \alpha \cdot \left( 1-\left| {{\theta }_{k}}-{{S}_{1}} \right| \right) \right\} \right]$=0,\\

simplified to get,\\

For ${i}$ ,by ${{\theta }_{k}}={{\theta }_{1}}$, \\

for ${{P}_{3}}\left[ {{P}_{4}}(\alpha +\beta )+\left( 1-{{P}_{4}} \right)(\alpha +\beta x) \right]+
\left( 1-{{P}_{3}} \right)\left[ {{P}_{4}} \right.\left. (\alpha +\beta \cdot (1-x))+\left( 1-{{P}_{4}} \right)\cdot \alpha  \right]=0$, \\

the following equation can be obtained in the same way:\\

For ${i}$, by ${{\theta }_{k}}={{\theta }_{2}}$, for ${{P}_{3}}\left[ {{P}_{4}}(-\alpha +\beta )+\left( 1-{{P}_{4}} \right)(-\alpha +\beta y) \right]+\left( 1-{{P}_{3}} \right)\left[ {{P}_{4}}( \right.-\alpha +\beta (1-y)+\left( 1-{{P}_{4}}(-\alpha ) \right]=0{}{}{}{}\left( Eq2 \right),$\\

For${j}$, by ${{\theta }_{R}}={{\theta }_{1}}$, for ${{P}_{1}}\left[ {{P}_{2}}(\alpha +\beta )+\left( 1-{{P}_{2}} \right)(\alpha +\beta x) \right]+\left( 1-{{P}_{1}} \right)\left[ {{P}_{2}} \right.(\alpha +\beta (1-x))+\left. \left( 1-{{P}_{2}} \right)\alpha  \right]=0$    ${(Eq3)}$\\

For ${j}$, by ${{\theta }_{k}}={{\theta }_{2}}$, for ${{P}_{1}}\left[ {{P}_{2}}(-\alpha +\beta )+\left( 1-{{P}_{2}} \right)(-\alpha +\beta y) \right]+$\\

$\left( 1-{{P}_{1}} \right)\left[ {{P}_{2}}(- \right.
$$\left. \alpha +\beta (1-y))+\left( 1-{{P}_{2}} \right)(-\alpha ) \right]=0$  
$(Eq4)$\\

Slove $\text{ }Eq1/2/3/4$ to get,\\

${{P}_{1}}={{P}_{3}}=(a/b)\left [  (x+y-2)/(x-y)\right ] $,\\

${{P}_{2}}={{P}_{4}}=(a/b)[(x+y)/(x-y)]$\\

\textbf{Comment1:}The result of Lemma 1 shows that the larger $a/b$ (that is, the greater the "risk-free utility obtained by individual participation" compared to the "risky utility obtained by joint participation"), the more individuals hope to choose to participate in collective action on a rational level; the analysis of $x/y$ and action probability can be carried out by drawing function images in simulation.\\

\textbf{Lemma2:}Under M1, when ${G}$ is $k-regularly$ connected ${graph}_{3}$ , I(t) = 0, ${P}_{tyi,n}(t)={P}_{tyi,n}$ and satisfy $\frac{1}{k}>\left( {{P}_{T{{y}_{i}},2}}/{{P}_{T{{y}_{i}},1}} \right)\cdot \underset{i}{\mathop{\max }}\,{{T}_{i}}(0)$, if $\exists i,{{a}_{i}}\left( 0 \right)=1$, then $t\to \infty $, $\forall i,{{a}_{i}}\left( t \right)=1$ , i.e. action synchronized with scale perception. In particular, if $k>\frac{1}{2}|V|$ , then, $t\to \infty $, ${{T}_{i}}\left( t \right)\to T$, threshold, action and scale perception are all synchronized.\\

\textbf{prove:}\\

{\large\textcircled{\footnotesize 1}}Threshold  $T(t)$ Synchronization\\

The introduction of random arrays, we can see thatand the/column of the line are 1,totraversing coefficients, and then the theorem of the random matrix can be seen.The neighboring domain must be overlapped, that is, letting goand Xing, there must be a column, then. Equivalence is established.\\

Thus,$\underset{i}{\mathop{\max }}\,{{T}_{i}}\left( t+1 \right)-\underset{i}{\mathop{\min }}\,{{T}_{i}}\left( t+1 \right)$\\

$d\left( {{v}_{i}} \right)>\frac{1}{2}|V|$$\text{r}(A)\left[ \underset{i}{\mathop{\max }}\,{{T}_{i}}(t)-\underset{i}{\mathop{\min }}\,{{T}_{i}}(t) \right]$\\

$=\left( 1-\frac{1}{k} \right)\left[ \underset{i}{\mathop{\max }}\,{{T}_{i}}(t)-\underset{i}{\mathop{\min }}\,{{T}_{i}}(t) \right]$\\
Further, by induction, we can get\\

$\underset{i}{\mathop{\max }}\,{{T}_{i}}(t)-\underset{i}{\mathop{\min }}\,{{T}_{i}}(t)$\\

${{\left( 1-\frac{1}{k} \right)}^{t}}\left[ \underset{i}{\mathop{\max }}\,{{T}_{i}}\left( 0 \right)-\underset{i}{\mathop{\min }}\,T_{j}^{{}}0 \right]$\\

Then $t\to \infty $, $\underset{i}{\mathop{\max }}\,{{T}_{i}}(t)-\underset{i}{\mathop{\min }}\,{{T}_{i}}(t)\to 0$\\

Then the thresholds $T\left( t \right)$ are synchronous, while $\underset{i}{\mathop{\max }}\,{{T}_{i}}\left( t \right)$ are clearly non-increasing.\\

{\large\textcircled{\footnotesize 2}}Scale perception P(t) is synchronized with action $a\left( t \right)$\\

Consider $P\left( t \right)$ , take the neighborhood $N\left( {{V}_{\text{k}}} \right)=\left\{ {{\text{v}}_{j}}:\exists i\in {{V}_{\text{k}}}{{\text{v}}_{j}}\in N(i) \right\}$ of ${{V}_{k}}\subseteq V$.\\

I. Need proof $\forall t\in N$:, make ${{V}_{k}}(t)=\left\{ i:{{a}_{i}}(t)=1 \right\},\forall j\in N\left( {{V}_{k}}(t) \right),{{a}_{j}}(t+1)=1$.\\

For this, we know from the condition that ${{V}_{k}}(t)$ is not empty, and\\

$N\left( {{V}_{k}}(t) \right)\subseteq N\left( {{V}_{k}}(t+1) \right)$.\\

For \\

$\forall j\in N\left( {{V}_{k}}(t) \right)$, $\exists i$, $j\in N(i)$,\\

then \\

$i\in N(j)$, \\

then \\

${{P}_{j}}(t+1)=\sum\limits_{N\left( j \right)}{{{a}_{k}}}(t)/\left| N\left( j \right) \right|1/|N(j)\mid =\frac{1}{\text{k}}$, \\

then \\

$\frac{1}{k}>\left( {{P}_{{{T}_{{{y}_{i,1}}}}}}/{{P}_{{{T}_{{{y}_{i,2}}}}}} \right)\cdot \underset{i}{\mathop{\max }}\,{{T}_{i}}(0)$, \\

then  \\

${{P}_{T{{y}_{j,1}}}}\cdot {{P}_{j}}(t+1){{P}_{T{{y}_{j,1}}}}\cdot \frac{1}{k}>{{P}_{T{{y}_{j,2}}}}\cdot \underset{i}{\mathop{\max }}\,{{T}_{i}}(0){{P}_{T{{y}_{j,2}}}}\cdot $\\

$\underset{i}{\mathop{\max }}\,{{T}_{i}}(t+1){{P}_{T{{y}_{j,2}}}}\cdot T_{j}^{{}}(t+1)$, \\

then \\

${{a}_{j}}(t+1)=1$ , \\

the conclusion is proved.\\

II. Further proof,$\forall l\in V$, $\exists t\in N,l\in N\left( {{V}_{k}}(t) \right)$.\\

In this regard, take]arbitrarily, recordas the distance between two points on, and be notified by, there is a path0]1]2], then3]4]and5]6]and7]8] , by induction,9]andare determined by. For all $l$, take $c=\text{max}d-1$, then $l\in N\left( {{V}_{k}}(c) \right)$, the proof is obtained.\\

According to II, $\forall l\in V,\exists t,l\in N\left( {{V}_{k}}(t) \right)$, and then I know ${{a}_{l}}(t+1)=1$, that is, action synchronization, then $t+2$ moments have proportional synchronization.\\

Combining {\large\textcircled{\footnotesize 1}}{\large\textcircled{\footnotesize 2}}, the threshold, action, and proportion are all synchronized, and the theorem is proved.\\

\textbf{Comment2:}The result of Lemma 2 shows that when the probability of individuals’ willingness to act on the rational level is high, and the social circle is medium-sized (k is greater than $1/2|V|$, but not too large), the system will generate more consistent views and collective actions.\\

\textbf{Theorem 1 (spontaneous action theorem):}Under M1, when \\

$k>\frac{1}{2}\left| V \right|$, \\

$\frac{a}{b}(x+y-2/x-y)>k\left[ 1-\frac{a}{b}(x+y-2/x-y) \right]$\\

and\\

$\frac{a}{b}[x+y/x-y]>k$$\left[ 1-\frac{a}{b}(x+y/x-y) \right]$,\\

the system \\

$(T(t),P(t),a(t))$ \\

is fully synchronized.\\

\textbf{prove:} \\

${{P}_{T{{y}_{1,1}}}}=(a/b)[(x+y-2)/(x-y)],{{P}_{T{{y}_{2,1}}}}=(a/b)[(x+y)/(x-y)]$ 
\\

obtained from Lemma 1, when $\frac{a}{b}(x+y-2/x-y)>k\left[ 1-\frac{a}{b}(x+y-2/x-y) \right]$ 
\\

and \\

$\frac{a}{b}[x+y/x-y]>k\left[ 1-\frac{a}{b}(x+y/x-y) \right]$ are satisfied, $\frac{1}{k}>\left( {{P}_{T{{y}_{i,2}}}}/{{P}_{T{{y}_{i,1}}}} \right)\cdot \underset{i}{\mathop{\max }}\,{{T}_{i}}(0)$, \\

and then obtained from Lemma 2 and \\

$k>\frac{1}{2}\left| V \right|$, \\

the system is completely synchronized.\\

\textbf{Remark3:}Theorem 1 is a direct corollary of Lemma 1/2, which establishes a relationship between private incentives/trust/circle size of individuals and collective action.\\

\textbf{Theorem 2 (action cycle theorem)} When it is a regular connected graph, and it satisfies, if there are and and, then there are ones that can be listed.\\

\textbf{prove:}\\

{\large\textcircled{\footnotesize 1}}Take the one that satisfies the conditions, by, according to the nature of the Brouwer degree. Further, from the monotonicity of \\

$\alpha >\max \left( {{k}_{1}},{{k}_{2}} \right)$ \\

and $f$, $f$ is subtracted from \\

$\left( \alpha ,T-\alpha  \right)\subset \left( {{k}_{1}},T-{{k}_{2}} \right)$, \\

then $\inf f\left( \alpha ,{{t}^{*}} \right)\frac{1}{c+1}$; \\

and according to the cycle, \\

$\inf f\left( mT+\alpha ,mT+{{t}^{*}}-{{\varepsilon }_{0}} \right)$$\frac{1}{c+1},{{\varepsilon }_{0}}$\\

is sufficiently small.\\

{\large\textcircled{\footnotesize 2}}When there is, and when, consider, from, we can know (Tup represents the update rule of the threshold T(t))\\

$\left\| T_{\text{u}p}^{t}\cdot T(0)-[\overline{T(0)}]_{i=1}^{n} \right\|$\\

$\le {{\left[ \sum{{{\left( \underset{i}{\mathop{\max }}\,T_{\text{u}p}^{t}T(0)[i]-T(0) \right)}^{2}}} \right]}^{\frac{1}{2}}}$\\

$\le {{\left[ \sum {{\left( \underset{i}{\mathop{\max }}\,T_{up}^{\text{t}}T(0)[i]-\underset{i}{\mathop{\min }}\,T_{up}^{\text{t}}T(0)[i] \right)}^{2}} \right]}^{\frac{1}{2}}}$\\

$\le {{\left[ \sum{{{\left( 1-\frac{1}{k} \right)}^{2t}}}{{\left( \underset{i}{\mathop{\max }}\,{{T}_{i}}(0)-\min {{T}_{i}}(0) \right)}^{2}} \right]}^{\frac{1}{2}}}$\\

$\to 0(t\to \infty )$\\

Then\\

$t\to \infty ,T(t)\overset{{{l}^{2}}}{\mathop{\to }}\,\left[ \overline{T(0)} \right]_{i=1}^{n},{{T}_{i}}(t)\to \overline{T(0)}$\\

Thus\\

$\forall \varepsilon >0,\exists t,\forall i,\left| {{T}_{i}}(t)-T(0) \right|<\varepsilon $, then further take 
\\

$\varepsilon =\frac{r}{2}/\left\{ \left[ \left( 1-f\left( t\bmod \left( T \right) \right)/f\left( t\bmod T \right) \right) \right]\cdot \left( {{P}_{TYi,2}}/{{P}_{TYi,1}} \right) \right\}$\\

$r=\frac{1}{k}-\left[ 1-f\left( t\bmod T \right)/f\left( t\bmod T \right) \right]\cdot \left( {{P}_{TYi,2}}/{{P}_{TYi,1}} \right)\overline{T(0)}$, 
\\

then $\exists T$ (determined) $t>T,T(t)-T(0)<\varepsilon $. 
\\

From the definitions of $f(t)>\frac{1}{c+1}$ 
\\

and \\

$c$, $\forall t>T$ 
\\

and \\

$\forall t>T,\frac{1}{k}>\left[ (1-f(t\bmod T)/f(t\bmod T)]\cdot \left( {{P}_{Ty,2}}/{{P}_{Ty,1}} \right)\overline{{{T}_{\left( 0 \right)}}} \right)$ 
\\

must have \\

$\frac{1}{k}>\left[ (1-f(t\bmod T))/f(t\bmod T) \right]\cdot \left( {{P}_{Ty,2}}/{{P}_{Ty,1}} \right)T(t)$. 
\\

Thus, if $\exists j\in N\left( i \right),{{a}_{j}}\left( t \right)=1$, 
\\

then $f(t\text{ mod }T)\cdot {{P}_{Ty,1}}{{P}_{i}}(t+1)f(t\text{ mod }T)\cdot {{P}_{Ty,1}},\frac{1}{k}>(1-f(t\bmod T))\cdot {{P}_{Ty,2}}\cdot T(t+1)$, 
\\

then ${{a}_{i}}\left( t+1 \right)=1$.\\

{\large\textcircled{\footnotesize 3}}It is known from the conditions,,, then, and then, it can be known from the idea of Lemma 2, for, fixed, then, and so on, and, it can be seen that there is a list that satisfies this condition, and the theorem is proved.\\

\textbf{Comment4:}Theorem 2 states that if the "exogenous cycle coefficient f(t)" and "exogenous event occurrence I(t) = 1" are coupled (that is, the event occurs in the rising stage of the cycle), and the length of the exogenous cycle allows sufficient diffusion of actions on the social network, then collective actions will emerge repeatedly.\\

\section{Simulation experiment and results}
\subsection{Simulation experiment design}
\subsubsection{Overview}
Since the above theorems are limited to relatively special situations, we hope to explore the conditions for achieving synchronization (large-scale collective action) and cycles (action fluctuations) in the model of this paper through numerical simulations, so as to answer questions 1 and 2—since the discussion of the latter (how the cycle is formed) depends on the answer to the former (the diffusion mechanism of a single collective action), we gradually carry out simulation experiments on the formation of collective action and the cycle of collective action. At the same time, in these two types of experiments, part of the experimental environment can be shared: (1) the small-world network is selected as the representation of social relations, which has a strong consistency with the network structure of the real world; (2) the expression of the willingness coefficient is set to ${f}_{i}(t)=\frac{1}{2}sin(m_{i,t}+n_{i})+\frac{1}{2}$, and ${m}_{i}=m$ and ${n}_{i}={n}$ can be used as individual homogeneity conditions.

\subsubsection{Formation of collective action}
According to the experimental environment described above, we can first consider the experimental design of "action formation". Starting from the variables, we can see that: as far as question 1-1 is concerned, it is mainly related to private utility $(\alpha)$ , public utility $(\beta)$, and type discrimination probability $(x/y)$ in the game model; as far as question 1-2 is concerned, we need to consider the distribution of individual types in the assimilation model (such as the proportion of type 1 is k) and the average degree of social networks $(d)$. In addition, since Question 1 focuses on the action formation/diffusion mechanism (rather than periodicity), we can set $f(t)=\frac{1}{2}$, $I(0)=(0…1…0)$ to eliminate the influence of exogenous system periodic variables.\\

In terms of experimental strategies, we can give a brief description of variable control and result visualization methods, which can be used as an example of "the impact of private utility $\alpha$ on collective action results", and experiments with other variables $(\beta/x/y/Ty/d)$ can be carried out according to this logic:\\
\begin{enumerate}[(1)]
\item Set different levels of variable , such as 0.1/0.3/0.5, and control other variables $\beta/x/y/Ty/d$ unchanged
\item Under a given level of $\alpha$ (such as $\alpha$=0.3), perform numerical simulation to obtain all the data in t updates of dynamics $(t\geq10)$.
\item Use the coloring network graph (such as "inactive individuals" is black, "action individuals" is red) to present the action status of all individuals at each moment, and present a diagram of all moments. As an example, if the dynamics are updated from time 0 to time 3, then under a specific parameter combination $(\alpha, \beta, x, y, Ty, d)$, we can obtain four dyed network diagrams $G(0)$, $G(1)$, $G(2)$, and $G(3)$, which represent the global individual states at four times respectively—different $\alpha$ obviously can have different diagrams.
\item Use the global action ratio  $Pro(t)=\sum \frac{a_{i}(t)}{N}$ to represent the proportion of the number of participants in the action at time t to the total population, and given time t (t=0...n), calculate the global action ratio under different $\alpha$ respectively. Finally, the relationship between "given $\alpha$, time t and global action ratio $Pro(t)$" is illustrated by a line graph.
\item For the parameter combination $(\alpha, \beta, x, y, Ty, d)$, the evolution result of individual threshold/proportion perception $({T}_{i}(t)/{P}_{i}(t))$ in the system can be presented selectively. In this regard, we can use the total deviation (Total Deviation) to describe the deviation degree of ${T}_{i}(t)$ from the consensus (the larger it is, the greater the deviation).
\end{enumerate}
Obviously, (1)/(2) is the experimental method, and (3)/(4) is the way to present the experimental results. According to (3) and (4), we compare the dyed network graph/line graph under different $\alpha$, so as to explore the influence of $\alpha$ on collective action in social networks. In the same way, this method can also be used to discuss $\beta/x/Ty/d$, thus constituting the experimental design of this article-in terms of calculation, we will use the networkx and numpy packages in python for code implementation. At the same time, we will use the two semantically equivalent expressions of "action synchronization" and "reaching collective action" (the former means that everyone takes the same action) to describe the experimental results.

\subsubsection{Collective action cycle}
The parameters in our paper are divided into two categories: one is structural parameters, which describe the features of the social network and its internal regions (namely communities); The second is individual parameters, which show the characteristics of individuals when integrating other people's opinions and carrying out migration behavior. These two types of parameters are represented by Greek letters in the above model, which is convenient for identification.\\

As far as structural parameters are concerned, we first set the type of graph G, which is a small-world network: to enhance the representativeness of our simulation, the experiment will include two cases, which respectively contain 50/500 nodes(to represent different population sizes), and the probability of random reconnection is 0.3. In addition, the simulated user migration behavior is limited to "duopoly competition mode," which means 'the number of communities is 2'. Finally, we set the noise/disturbance in the evolution of opinions, which can be randomly selected between the minimum and maximum values of opinions at the time of t and accords with the "uniform distribution" mentioned above. \\

In terms of individual parameters, we set different confidence thresholds d in model 1, including 1, 0.8, and 0.3, where 1 is an extreme case, indicating that individuals in the network can influence each other's opinions at any time. At the same time, we set parameters $\varphi$ and $\sigma$: the former includes 1, 0.5, 0.4, 0.09, etc. (1-$\varphi$ represents the degree of individual being affected by noise), while the latter has values of 1, 0.9, 0.4, etc. (representing the reduction rate from private opinions to expressed opinions, which can be regarded as the effect of "the spiral of silence"). Finally, we consider the parameters $\delta$ when individuals make migration decisions, and $\delta$/1-$\delta$ represent the influence of social distance/opinion distance, respectively. Therefore, we set them to 0.8 and 0.3, representing individuals with the intimacy tendency and opinion convergence tendency in current social media. 

\subsection{Simulation results and analysis}
\subsubsection{Formation of collective action}
According to the above strategies, we explore the influence of variables in the game model and the assimilation model on the action results to answer questions 1 and 2 respectively. In the game model, $\alpha$ and $\beta$ are obviously directly related to the “intentional action probability” of labor activists/active activists: when $x/y$ is given and within a valid range, $\alpha$/$\beta$ is positively correlated with the latter—this is reflected in the function graph in Figure 1. The explanation of this property can be very intuitive: since $\alpha$ represents “utility that active actors must obtain after participation” (independent of others’ choices), and $\beta$ represents “utility that can be obtained through joint participation” (depending on others’ choices), the former is less risky than the latter. When $\alpha$/$\beta$ is larger, actors can obtain higher utility at a lower risk level, and thus are willing to participate in the action.\\

In addition, by referring to Figures 1.2 and 1.3, we can also find that the rise of $\alpha$/$\beta$ has a positive effect on the synchronization result and speed (whether the collective action is achieved/the length of time to achieve it). For example, the blue/yellow/green curves all show synchronization and faster speed. The resources obtained (there is a risk incentive that the other party does not participate), and the utility that can be promoted only by the individual's own participation (risk-free incentive) will have a stronger positive impact on the "willingness to act". This mechanism can be applied to restrictive measures to collective protest, such as reducing the no/low-risk utility (salary) that can be directly obtained by wage-seeking protest.\\

Figure 1: 1.1-1.3 represent "the evolution of the proportion of collective action under different $\alpha$/$\beta$" and "the functional relationship between the probability of willing action/mixed equilibrium probability Pty1 and $\alpha$/$\beta$", and the values of other parameters are x = 0.2/y = 0.9/d = 5/k = 1

\begin{figure}[htbp]
	\centering
	\begin{subfigure}{0.32\linewidth}
		\centering
		\includegraphics[width=0.9\linewidth]{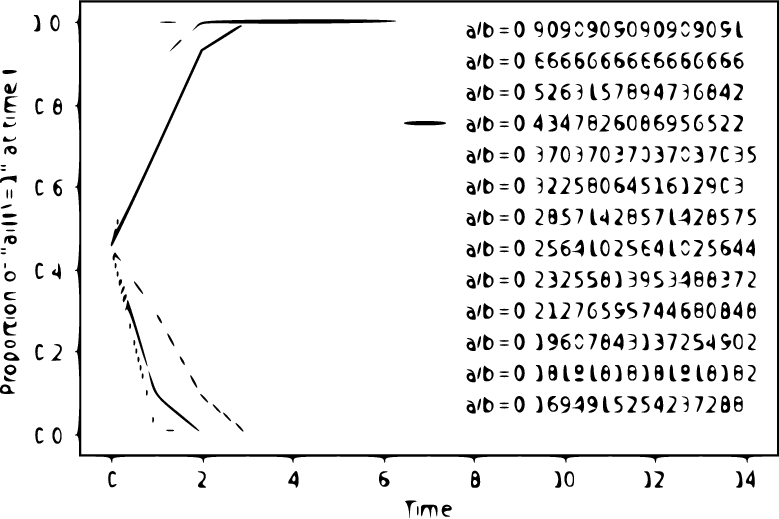}
		\label{chutian3}
	\end{subfigure}
	\centering
	\begin{subfigure}{0.32\linewidth}
		\centering
		\includegraphics[width=0.9\linewidth]{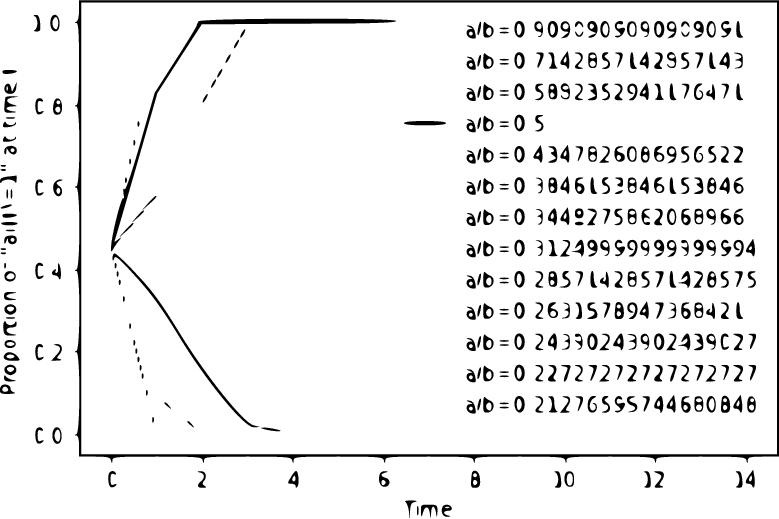}
		\label{chutian3}
	\end{subfigure}
	\centering
	\begin{subfigure}{0.32\linewidth}
		\centering
		\includegraphics[width=0.9\linewidth]{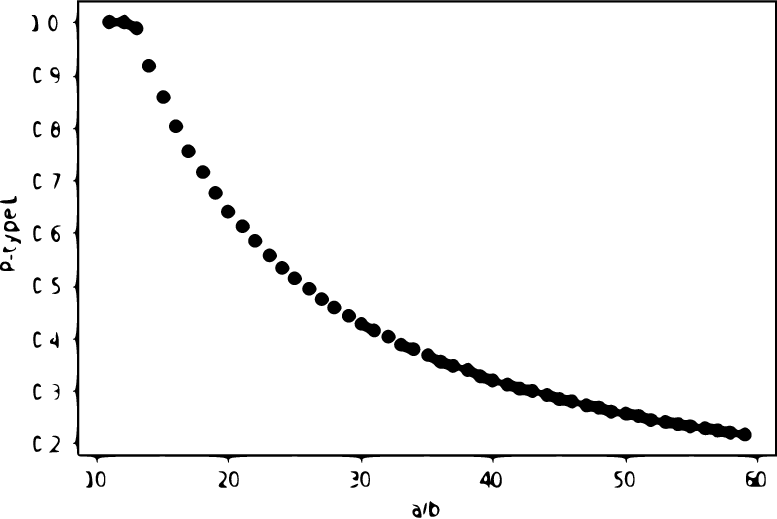}
		\label{chutian3}
	\end{subfigure}
	\caption{}
	\label{da_chutian}
\end{figure}

As for the parameter $x/y$, we found through simulation that when $x/y$ is large, global synchronization is easy to achieve and the speed is faster. Since $x$ represents the “probability that an active person i believes that individual j is an active person”, and $y$ represents the “probability that a non-active person believes that individual i is an active person”, the simulation results may come from a mechanism: when an activist/labor movement person agrees that his peers have a tendency to act actively/struggle ($x$ is high), he will simultaneously obtain the utility of $\alpha$ + $\beta$ after taking action (which is far beyond the zero utility caused by inaction, thus increasing the willingness to participate in the action of the active person); (It may even be smaller than the zero utility caused by "inaction", so it may not necessarily increase the willingness to act of the non-active) - In addition, there is a 2 $\alpha$ gap in utility between the former and the latter, which comes from the cooperation and private reporting of similar individuals. Therefore, when $x/y$ is large, the willingness to act of non-active individuals may not be affected, but the willingness to act of activists/labor activists may be greatly increased to facilitate the achievement of collective action. This conclusion can be combined with the discussion on the two types of incentives in the previous paragraph to form an answer to Question 1. Since this result takes into account the interaction and beliefs between individuals, it is closely related to the empirical research on "communicative relationship/trust among workers", which can further induce the assimilation and social network explanation of collective action below.\\

figure 2: Both 2-1 and 2-2 reflect the influence of the ratio of "prior cognition $x$" (the probability that the active actor thinks the other party is positive) to "prior cognition $y$" (the probability that the non-active person thinks the other party is positive) on synchronization, and the other parameters are a=0.1/b = 0.15/d = 5/k = 1.\\
\begin{figure}[htbp]
	\centering
	\begin{subfigure}{0.45\linewidth}
		\centering
		\includegraphics[width=0.9\linewidth]{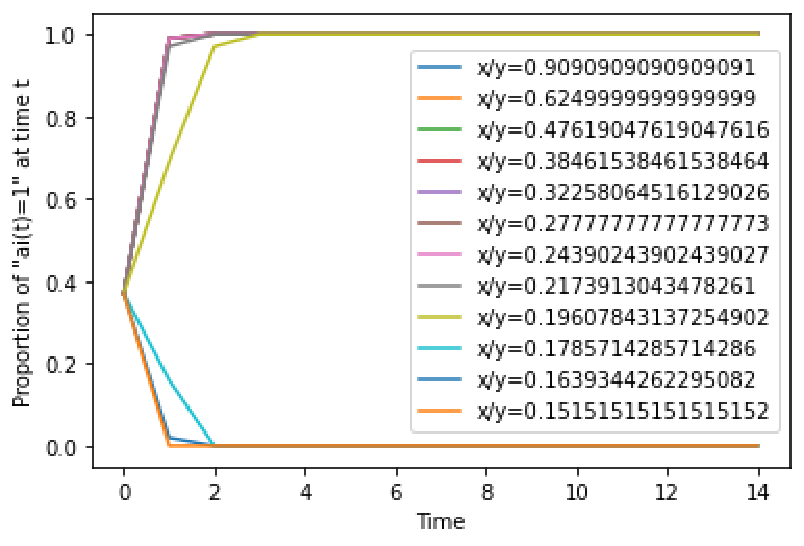}
		\label{chutian3}
	\end{subfigure}
	\centering
	\begin{subfigure}{0.45\linewidth}
		\centering
		\includegraphics[width=0.9\linewidth]{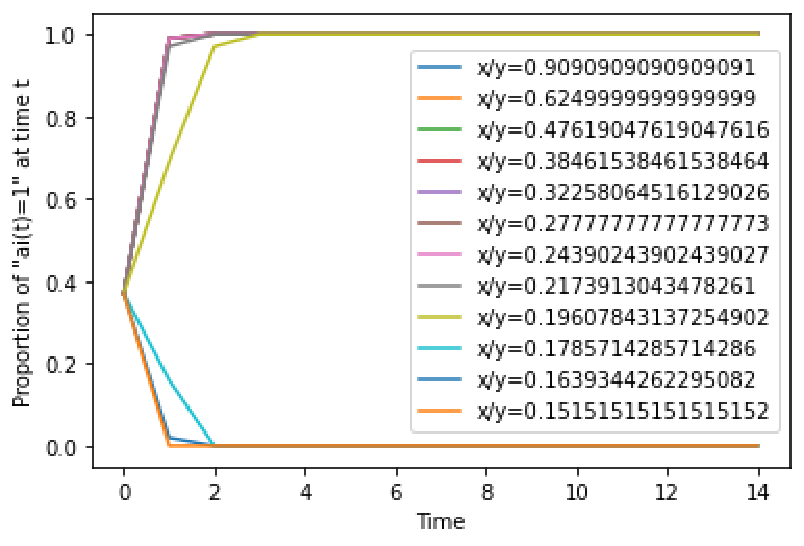}
		\label{chutian3}
	\end{subfigure}
	\caption{}
	\label{da_chutian}
\end{figure}

In the assimilation model, k (type 1/proportion of active actors) and d (average network degree) can be discussed separately. From Figures 3-1 and 3-2, and Figures 3.e and 3.f, it can be seen that when the proportion of k is large, the model is easy to achieve global synchronization, and the rate of joint action is faster. The light blue curves in both figures can reflect this conclusion—this conclusion is consistent with our theoretical intuition, that is, "when the proportion of activists is large, there is a higher possibility of large-scale collective actions in the system." At the same time, it shows that the nature of the game is effectively inherited under the dynamics framework. This is because activists/labor activists tend to choose this behavior because they get higher benefits when they fight. In addition, we also observed the "resistance/threshold dynamics" at different k times (that is, the evolution of cognitive resistance Ti(t) over time), and found that consensus can be reached in a relatively short period of time, which is closely related to the "small group nature" of the social network selected in this paper.
image 3: 3-a and 3-b describe the evolution of the system’s “proportion of actors” over time under different types of proportions k (proportion of active actors); Figures 3-c and 3-d show the evolution process of the gap/deviation between the individual threshold/resistance $T_{i}(t)$ and the synchronization state (consensus) over time; Figures 3-e and 3-f visually show the arousal of action when k=0.4 (yellow/black indicates action/inaction); in addition, $\alpha$ = 0.7, beta = 0.3, x= 0.8, y=0.9, d=3.\\

\begin{figure}[htbp]
	\centering
	\begin{subfigure}{0.325\linewidth}
		\centering
		\includegraphics[width=0.9\linewidth]{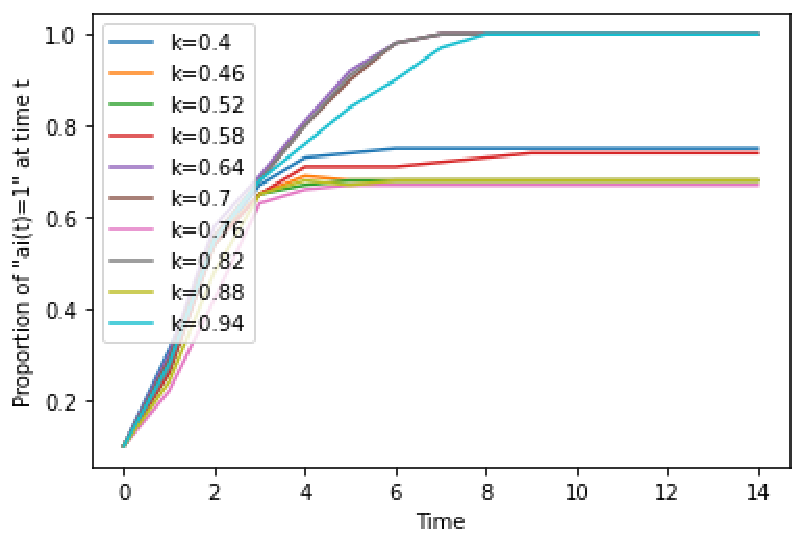}
		\caption{chutian3}
		\label{chutian3}
	\end{subfigure}
	\centering
	\begin{subfigure}{0.325\linewidth}
		\centering
		\includegraphics[width=0.9\linewidth]{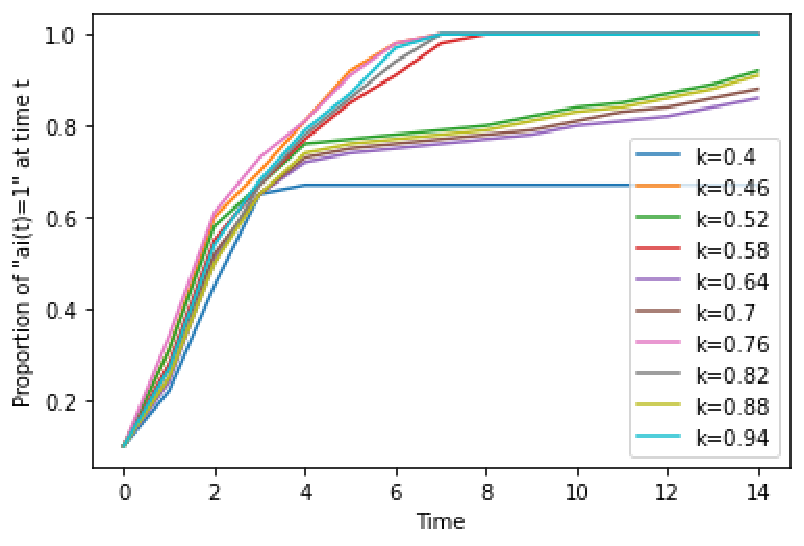}
		\caption{chutian3}
		\label{chutian3}
	\end{subfigure}
	\centering
 
	\begin{subfigure}{0.325\linewidth}
		\centering
		\includegraphics[width=0.9\linewidth]{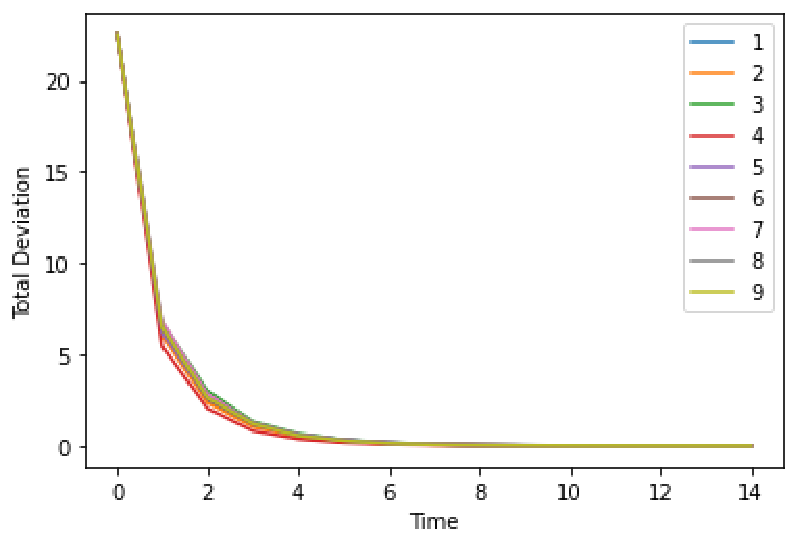}
		\caption{chutian3}
		\label{chutian3}
	\end{subfigure}
 	\centering
	\begin{subfigure}{0.325\linewidth}
		\centering
		\includegraphics[width=0.9\linewidth]{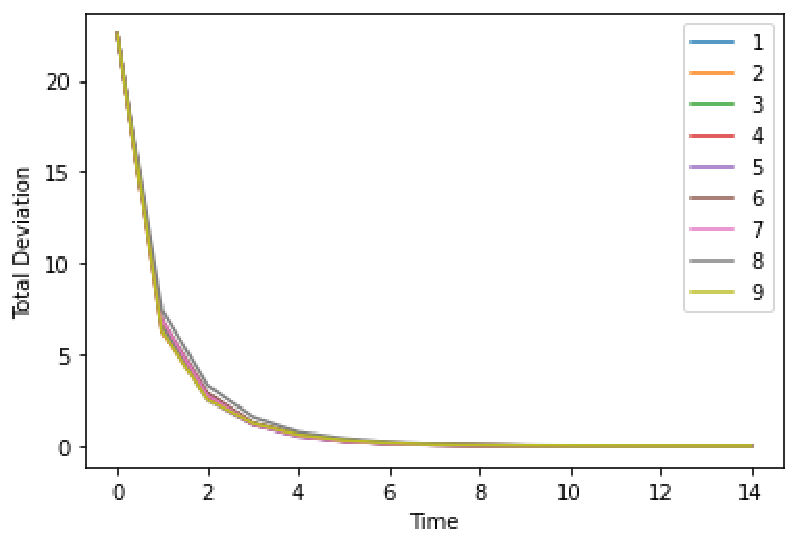}
		\caption{chutian3}
		\label{chutian3}
	\end{subfigure}
	\centering
 
	\begin{subfigure}{0.325\linewidth}
		\centering
		\includegraphics[width=0.9\linewidth]{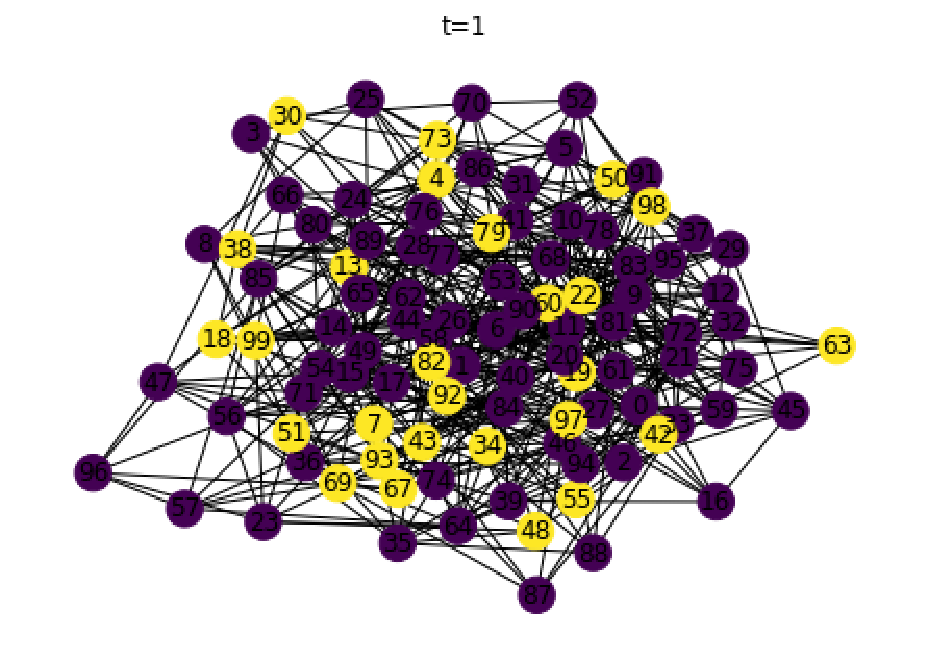}
		\caption{chutian3}
		\label{chutian3}
	\end{subfigure}
	\centering
	\begin{subfigure}{0.325\linewidth}
		\centering
		\includegraphics[width=0.9\linewidth]{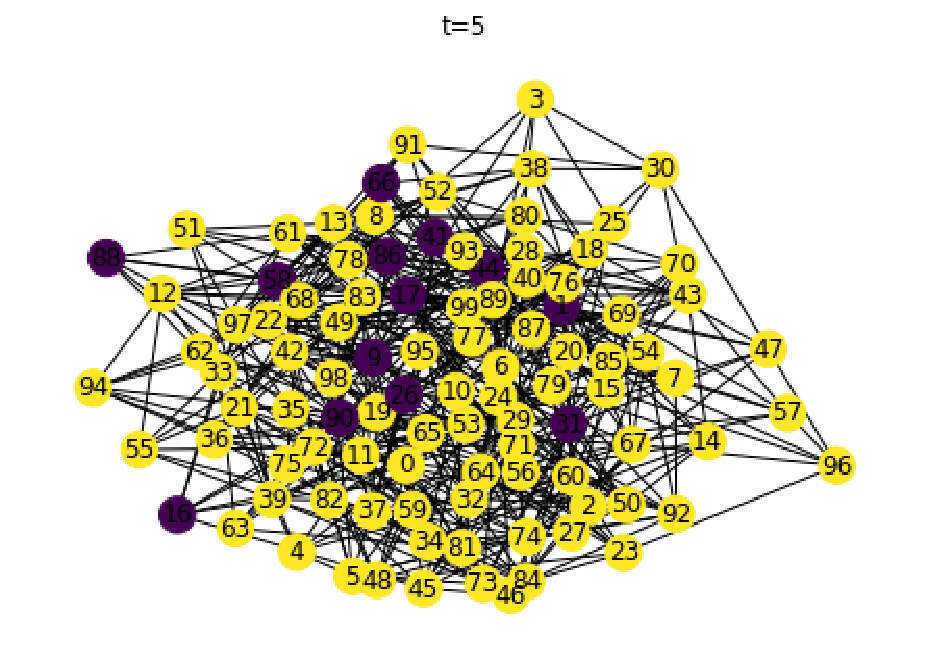}
		\caption{chutian3}
		\label{chutian3}
	\end{subfigure}
        \caption{}
	\label{da_chutian}
\end{figure}

As far as d is concerned, it can measure the impact and scope of individual actions/thresholds: According to Figures 4-1 and 4-2, under the condition of given collective action (action synchronization) parameter combination $(\alpha, \beta, x, k_{i})$, the average degree d and action synchronization rate/synchronization possibility are not linear, but obey a "central" curve: in the range of d greater than 0 and less than 10, if d is extremely high/low (such as d$\geq$8/d$\leq$2, the gray/orange curve in the left/right figure), collective action often It cannot be achieved; if d is close to the middle value (such as d=5/d=6, the purple/brown line in the left figure), the action can not only move towards synchronization, but also the speed is better than the results under the parameters at both ends.\\

A reasonable explanation for this is: from the perspective of the overall average, the value of d represents the social scope of the individual. When d expands/shrinks, the number of neighbors that affect individual resistance (threshold T)/thrust (proportion P) also increases/decreases—if the number of individual neighbors is insufficient (such as d=2), the "participating actors" cannot establish relationships with many individuals, and it is difficult to trigger collective actions; If the proportion is relatively large, the resistance suffered by individuals is often greater than the thrust, so they also refuse to participate in collective actions; only when the number of individual neighbors is moderate (such as d=5), the "participating actors" can first promote small-scale local synchronization (such as 3 interrelated people/3-element closures all participate in actions), and gradually persuade global individuals to join their ranks to achieve collective action. This conclusion can be combined with the discussion on the proportion of active activists ki in the previous paragraph as the answer to question 2. In addition, the growth of d allows us to observe a typical "phase transition" phenomenon: given the ratio of $(\alpha, \beta, x, {k}_{i})$ to the initial actor, there is a certain "critical point" determined by the average degree d of the network, and its value is recorded as d*. When d $\textgreater$ d* or d$\textless$d*, the limit state of the system is "all people achieve collective action" (global synchronization of $a_{i}(t))$; otherwise, if d = d*, only a few people/no one in the system will take action.\\

Figure 4: Among them, Figures 4.a and 4.b describe the evolution of the "proportion of actors involved" in the system over time under different average degrees d; Figures 4.c-4.f describe the action status of nodes at different moments when d=3 (yellow/black means action/non-action), and $\alpha$ = 0.7, $\beta$ = 0.3, x=0.8, y=0.9, k=0.6.\\

\begin{figure}[htbp]
	\centering
	\begin{subfigure}{0.325\linewidth}
		\centering
		\includegraphics[width=0.9\linewidth]{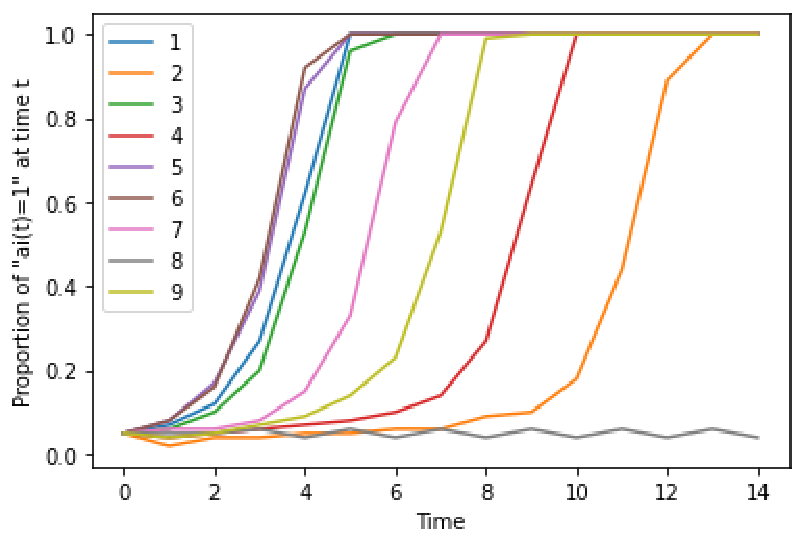}
		\caption{chutian3}
		\label{chutian3}
	\end{subfigure}
	\centering
	\begin{subfigure}{0.325\linewidth}
		\centering
		\includegraphics[width=0.9\linewidth]{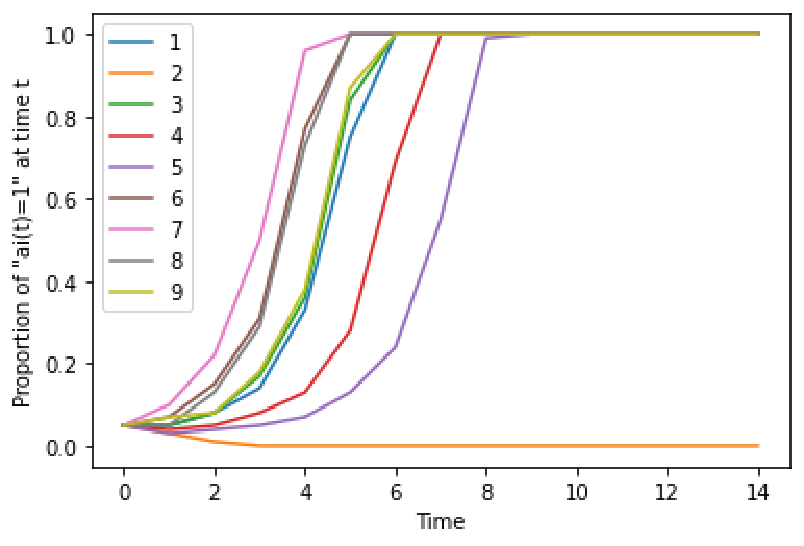}
		\caption{chutian3}
		\label{chutian3}
	\end{subfigure}
	\centering
 
	\begin{subfigure}{0.325\linewidth}
		\centering
		\includegraphics[width=0.9\linewidth]{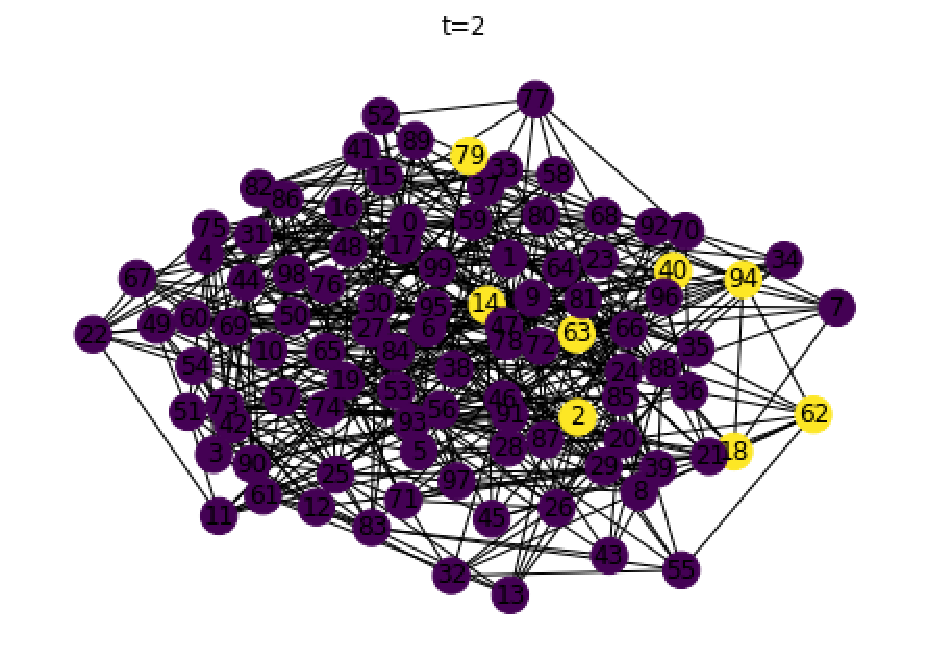}
		\caption{chutian3}
		\label{chutian3}
	\end{subfigure}
 	\centering
	\begin{subfigure}{0.325\linewidth}
		\centering
		\includegraphics[width=0.9\linewidth]{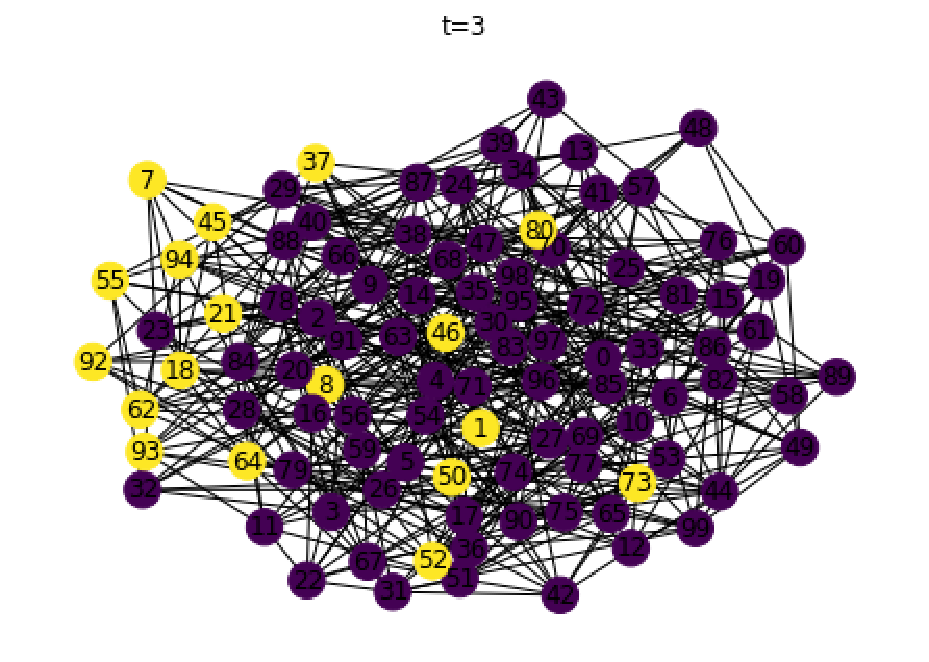}
		\caption{chutian3}
		\label{chutian3}
	\end{subfigure}
	\centering
 
	\begin{subfigure}{0.325\linewidth}
		\centering
		\includegraphics[width=0.9\linewidth]{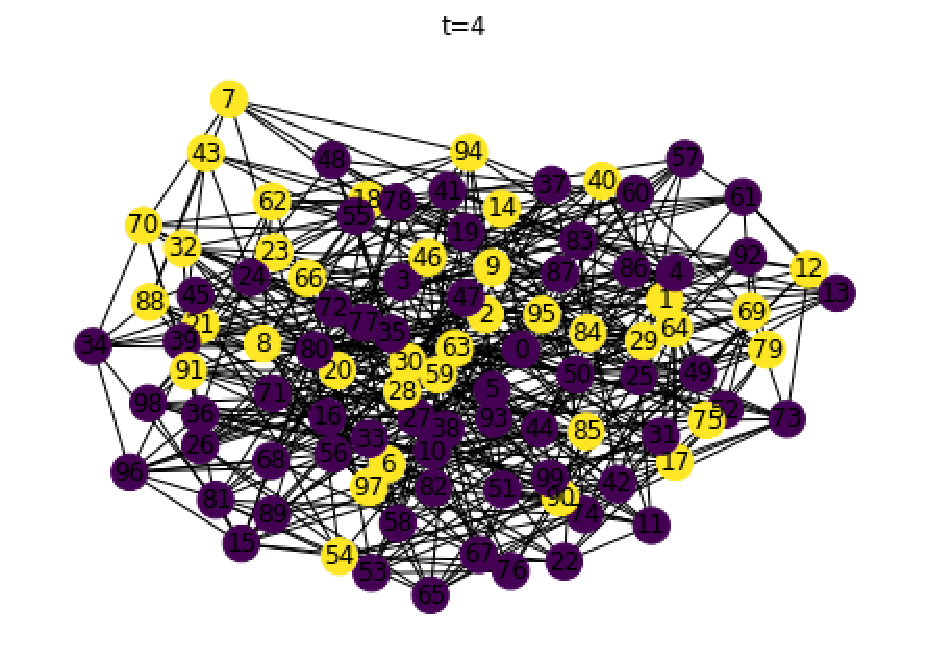}
		\caption{chutian3}
		\label{chutian3}
	\end{subfigure}
	\centering
	\begin{subfigure}{0.325\linewidth}
		\centering
		\includegraphics[width=0.9\linewidth]{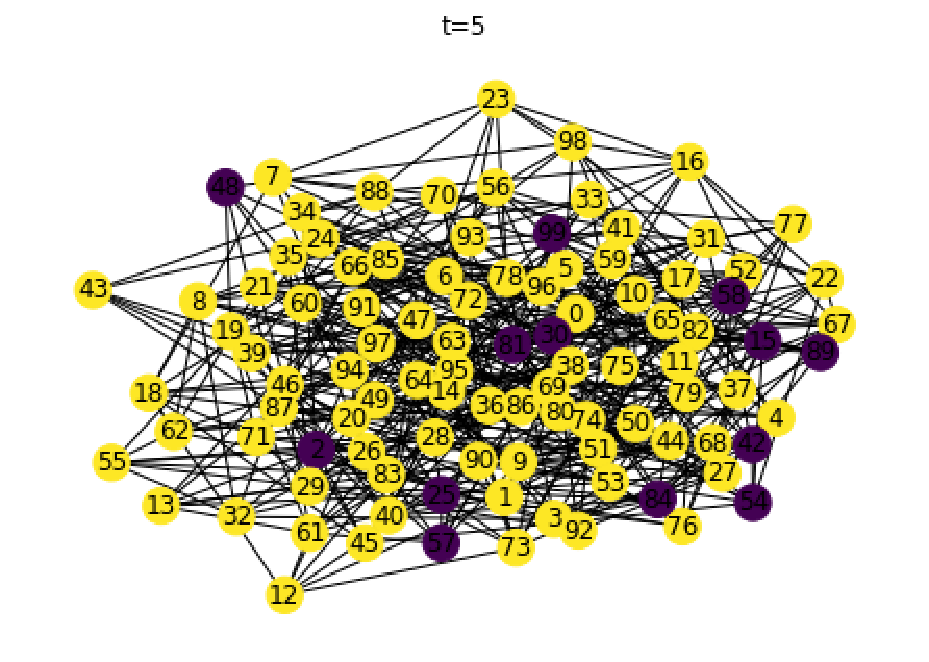}
		\caption{chutian3}
		\label{chutian3}
	\end{subfigure}
        \caption{}
	\label{da_chutian}
\end{figure}

In sum, the number and strength of ties between individuals (identified as social capital under certain conditions) will have a non-linear effect on collective action and thus serve as an important mechanism for enabling/truncating worker resistance. At the same time, in the above simulation process, we can observe an important phenomenon, that is, the emergence of "clusters" - this is reflected in multiple results such as Figure 3-e/3-f, Figure 4.c-4.f, the yellow nodes that choose to act often form small groups and gradually spread outward. In this regard, we can also establish an explanation based on theoretical assumptions: in the game equilibrium of individuals, if the probability Ptyi of the willingness to act is larger, then under the rules of dynamic evolution, a small number of actors at time t can drive most individuals in their neighborhood to form a local synchronization/small cluster. This phenomenon can explain a kind of "trigger-activation" mechanism, that is, for individual i, under the condition of "the behavior is risky but the expected return is high", if the surrounding individual j adopts this behavior, individual i only needs a weak imitation orientation to combine the information from j with his own will (trigger), and also adopt this behavior (activation), resulting in the emergence of action clusters. In addition to the above conclusions on the impact of specific variables, this "trigger-activation" mechanism should also be included in the important considerations for understanding collective action.

\subsubsection{Collective action cycle}
According to the method in 3.1.3, we further simulate the above model, and separately discuss the influence of "periodic coordination degree" and "effective period length" on the periodicity of collective action. In terms of cycle coordination, Figure 5 presents the basic results. When Dev(T1, T2) is low, the exogenous decay coefficient f(t) and the exogenous activation I(t) fluctuate more closely over time, thus giving rise to multiple collective actions with a higher proportion of participants—in 5-2, the blue/yellow line in the last cycle is like this (the difference is 0/1, respectively). In addition, the difference between T1 and T2 (which also reflects the size of Dev) can also lead to differences in the duration of collective action: when the difference between the two is small, the duration of collective action tends to be longer (the brown line in Figure 5-1). At the same time, we can also find in 5-1: Whether the initial action in a specific cycle can be aroused is not directly related to the length of the cycle, which leads us to the next question, namely the effect of "effective cycle length".\\

Figure 5: Both 5-1 and 5-2 describe\\
\begin{figure}[htbp]
	\centering
	\begin{subfigure}{0.45\linewidth}
		\centering
		\includegraphics[width=0.9\linewidth]{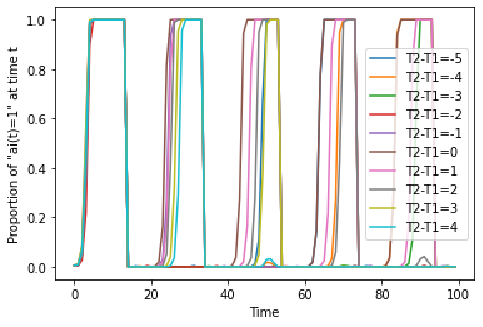}
		\label{chutian3}
	\end{subfigure}
	\centering
	\begin{subfigure}{0.45\linewidth}
		\centering
		\includegraphics[width=0.9\linewidth]{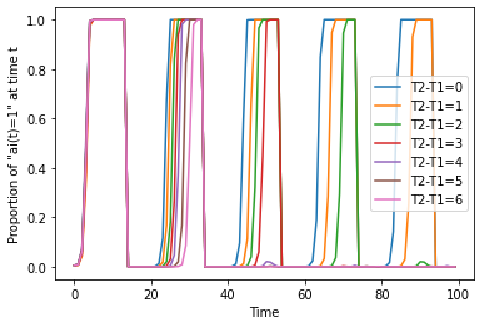}
		\label{chutian3}
	\end{subfigure}
	\caption{}
	\label{da_chutian}
\end{figure}

In terms of the length of the effective cycle (denoted as T), the simulation results in the following picture show: given the parameter conditions such as $\alpha$/$\beta$/d, when T reaches a certain value, the model will undergo a "phase transition" - when T<8 (such as the case of T=6/yellow line), because the period of time when the individual has a high willingness coefficient in each cycle is too short, the system is often difficult to achieve synchronization, so only a certain proportion of people participate in collective actions; when T$\geq$8 (such as the green/red line), a longer effective cycle allows individuals to be aroused/participated by people around Actions to facilitate the diffusion of collective actions and achieve synchronization in the overall social network. At the same time, the value of T that meets this condition does not affect the probability of synchronization in a long period of time, but only involves the time and rate required for synchronization.\\
\begin{figure}[htbp]
	\centering
	\begin{subfigure}{0.45\linewidth}
		\centering
		\includegraphics[width=0.9\linewidth]{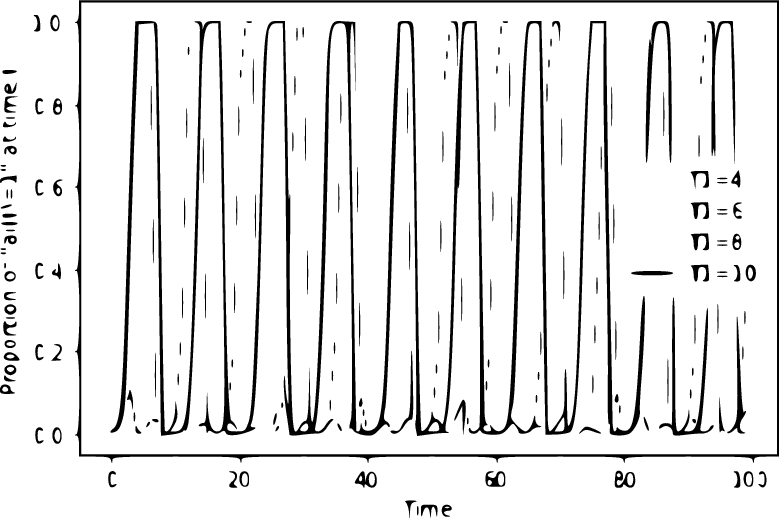}
		\label{chutian3}
	\end{subfigure}
	\centering
	\begin{subfigure}{0.45\linewidth}
		\centering
		\includegraphics[width=0.9\linewidth]{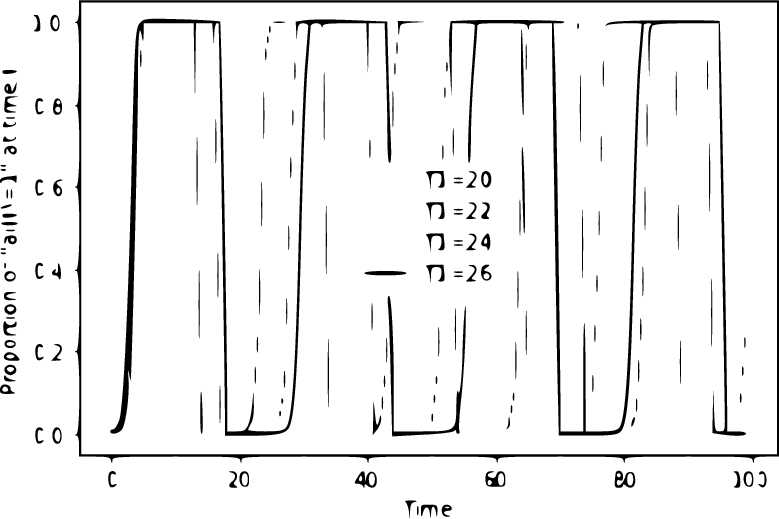}
		\label{chutian3}
	\end{subfigure}
	\caption{}
	\label{da_chutian}
\end{figure}

Finally, it is worth noting that the above experiments were all carried out under the conditions of “formation of collective action”, such as large $\alpha$/$\beta$, middle d, etc. Such parameters can be combined with the exogenous willingness coefficient f(t)/exogenous activation I(t) to make individual willingness fluctuate and effectively promote the diffusion of collective action when it is near the peak—at the same time, parameters such as $\alpha$, $\beta$, and d have a limiting effect on the maximum scale of collective action (if the conditions are not met, it may lead to the failure of collective action), so they can be used as a necessary condition for the generation of a collective action cycle.\\

\section{Empirical Data Calibration}
\subsection{Data and methods}
The model calibration data in this paper contains a number of macro variables, namely: "number of people", "region", "time" and "industry" in the transportation industry, which come from the "Map of Collective Actions of Chinese Workers", which can be obtained from the \href{https://baike.baidu.com/item/LaTeX/1212106}{webpage}etc. for public access. For this data set, we choose the collective actions of the express delivery industry as the core object, and remove the labor protest movements that have suffered from government intervention, so as to satisfy the "spontaneity" feature stipulated by the model in this paper (only applicable to spontaneous collective actions).
\\
Based on this, we use this time series data to estimate the kinetic model. Based on the above data set, the estimation method used in this paper is mainly based on the kernel density method and Bayesian calculation. The idea is: (1) According to the approximate Bayesian calculation, sample the parameters ($\alpha$, $\beta$, x, Ty, d) to be estimated; (2) According to the sampled values, use Kernel Density Estimation (KDE) to fit the posterior density; (3) Calculate the characteristics of the density function obtained by the MCMC method, such as expectation/Bayesian factor, etc., so as to obtain the posterior expectation estimation of the parameters (see my working for specific methods paper\cite{19}). On the basis of parameter estimation, we can calculate the difference between the predicted value and the real value to judge the fitting degree of the model (due to the limitation of time and computer computing power/cannot do high-dimensional identical distribution test of data like the working paper).

\subsection{Analysis of statistical results}
Based on the above methods, we estimated the values of the eight parameters $\alpha$/$\beta$/x/y/ap/bp/aT/bT respectively, which are 0.569, 0.739, 0.52, 0.934, 0.532, 0.476, 0.452, and 0.525, which can be used to illustrate the nature of this case (labor movement). It can be seen that in the labor movement, the proportions of “private utility” and “public utility” are relatively close, and non-activists/labor activists are twice as likely to view others as activists than activists. At the same time, we evaluate the fitting effect of the model, which can be presented in the figure below. It can be seen from Figure 5 that the overall trend of real data and simulated data is relatively close to illustrate the reliability of the model.\\

Based on this, we can further explore the practical significance of the parameter reflection. Obviously, the ratio of $\alpha$/$\beta$ is less than 1, indicating that it may be difficult for protest participants to obtain effective “risk-free incentives.” This may be due to the extremely high personal and economic risks of this behavior (facing imprisonment and fines), and it is often difficult to successfully obtain private rewards during the protest (such as successfully asking for salary), so it cannot lead to higher inner benefits. For x/y, this parameter is about 0.5, which means that during the course of action, activists/labor activists have insufficient belief in the movement tendency of others, and cannot reach effective cooperation, such as large-scale joint resistance. Finally, ap/bp/aT/bT are all positive, and the distribution is relatively concentrated, which can verify the functional relationship between the "threshold"/"proportion" and "action resistance"/"action thrust" in the theory to illustrate the reliability of the mechanism.\\

Figure 5: The fit, where yellow indicates the number of collective actions per month in the real data and blue indicates the number obtained by the simulated model.\\

\begin{figure}[htbp]
	\centering
	\includegraphics[width=0.9\linewidth]{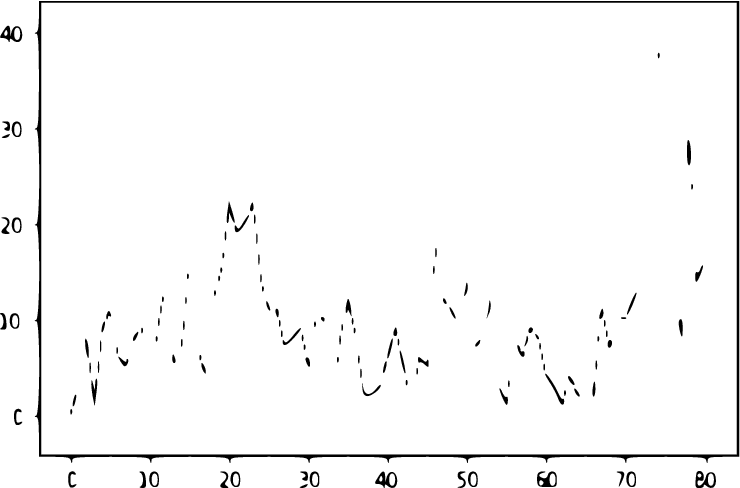}
	\caption{}
	\label{Fig:**}
\end{figure}

\section{Theoretical Construction: Spontaneous Synergy Theory}
\subsection{Overview}
Based on the above theoretical proof, numerical simulation and empirical analysis, we can put forward a general theory of spontaneous collective action. For this, we first need to clarify the meaning of the word "spontaneity": it means that the individual action state (action or non-action) in the system is not subject to direct external control, that is, "there is no mandatory rule that makes $ai(t)$ = 0 or $a_{i}(t)$ = 1 when attributes such as Ti(t) and Pi(t) do not meet the appropriate conditions". Obviously, the direct control of $a_{i}(t)$=0/$a_{i}(t)$=1 presents two types of means, the former is the suppression of collective action (such as the restriction of personal freedom of protesters, etc.), and the latter is a way of forcing participation (such as the requirements of unit activities, etc.).\\

Excluding such controls, we theorize cycles for collective action that may be termed "spontaneous cycles" to answer both questions 1 and 2. For question 1, we can give a basic framework from two levels, namely "formation conditions" and "diffusion mode", respectively serving 1-1 and 1-2: What is the trigger method of specific collective actions? How does it relate to both rational decision-making and social assimilation? This part can be called the theory of spontaneity, focusing on the endogenous diffusion of collective action after it has been initialized/evoked (i.e., an event in which initially very few individuals start acting).\\

For question 2, we can consider the conditions and generation mechanism behind the fluctuation of collective action, which involves the relationship between the fluctuation of exogenous will f(t), exogenous event I(t) and the endogenous system (that is, the evolution of T(t)/P(t) in the social network). It is worth noting that the phenomenon revealed by the spontaneity theory (short-term specific collective action) satisfies special periodic conditions, while the results of the cycle theory (repeated fluctuations of collective action) depend on the general mechanism of the spontaneity theory.

\subsection{Theory of Spontaneity: Conditions and Mechanisms}
\subsubsection{Formation conditions}
As far as the formation conditions are concerned, we found four important parameters that promote collective action in the proof and simulation: $\alpha$/$\beta$, x/y, d, k, which respectively represent the "ratio of private (action) utility to common (action) utility", "ratio of active actors agreeing with each other's active actions/non-active actors agreeing with each other's active actions", "average number of individual connections in social networks", and "proportion of active actors". Overall, we can name these four parameters as four theoretical concepts: “Risk-Free Incentive Ratio” (RP), “Cooperative Belief Ratio” (BP), “Average Social Ties” (ST), and “Activity Level” (AP).\\

The naming of these four concepts is closely related to the previous analysis: RP is because $\alpha$/$\beta$ represent risk-free income/risky income (the acquisition is independent of/depending on the actions of others); BP represents the degree of cooperation between individuals (high x means that active individuals of the same type can recognize each other/jointly participate in actions); ST can explain the average number of individuals affected/influenced in the network (high d means close connections facilitated by a large number of social ties); Based on this, we found a sufficient set of elements to trigger collective action:
\begin{enumerate}[(1)]
\item The risk-free incentive ratio is higher than $\alpha$/$\beta$.
\item Cooperative belief is higher than x/y.
\item The average amount of social ties d (social circle size) is in the middle.
\item The degree of activity k is high.
\end{enumerate}
In Part A, aiming at (1), it expands Olson’s proposition of “selective incentives/collective incentives”, trying to explain that the degree of risk involved in incentives (the probability of gaining benefits) is an important factor in achieving actions, and individuals have a strong sensitivity to “only requiring their own participation/not requiring others to cooperate and achieve collective action benefits”. As for (2), it reflects the influence of cooperative beliefs on actions, which may reduce individuals’ risk perception of risk incentives/"cooperative benefits" (requiring per capita action to obtain), thus serving as a supplement to (1). Therefore, conditions (1) and (2) of part A can be regarded as "rational decision-making conditions" for collective action.\\

In Part B, for (3), this condition can refute a common myth from the social network school, that is, “individual social ties enhance social capital/social capital leads to action participation”, which is because social ties not only belong to an individual’s “resource” (such as available social capital), but also exist as its “source of influence” (convince/oppose him/her to participate in collective action); therefore, at the early stage/when there are only initial actors, too many/too few neighbors will hinder the individual’s participation in the action—the former leads to the proportion perception that promotes action Insufficient (influenced), the latter makes it difficult for existing actors to promote the actions of others (influence), which can be jokingly called the mechanism of "shooting the first bird" (M)/"slapping the bird with a slap" (N). Regarding (4), when most individuals have a high willingness to act, it can effectively inhibit the two mechanisms of M/N as a supplement to (3). Thus, conditions (3) and (4) of part B can be regarded as the "conditions of social assimilation" for collective action.\\

All in all, we integrate the framework of rational decision-making and social assimilation through conditions (1) (2) (3) (4), which may bridge the understanding of collective action in public economics and social dynamics, and make corresponding supplements and clarifications to the classic views—such as the insights of Olson/social network school, so as to form a more effective theoretical explanation called the "spontaneous collective action" framework.

\subsubsection{Diffusion mechanism}
As far as action diffusion is concerned, one micro-mechanism and another macro-phenomenon can be paid attention to at the same time—the former is a “trigger-(endogenous) activation” process (different from the exogenous activation described later), and the latter is a “cluster” feature. Both are presented in simulations, and the following attempts to detail their elements in order to facilitate the rigor of the theory.\\

"Trigger-(endogenous) activation" constitutes the same process in the dyeing network above, that is, "nodes change from black to yellow" (from inaction to action), and it is difficult to effectively separate the two words; therefore, they should be discussed separately here. On the level of triggering, individual i’s threshold/resistance Ti(t) and proportional cognition/thrust Pi(t) are simultaneously triggered and updated by the neighbors who affect him, which is the cognitive/attitudinal response caused by “social assimilation”; on the level of endogenous activation, the individual needs to refer to the “degree of willingness to act based on his own utility Ptyi” and compare the magnitude of resistance and thrust to choose whether to act, which is the result of “rational decision-making”.\\

Based on the above discussion, the “trigger-(endogenous) activation” mechanism is an explanation of “how the two aforementioned formation conditions (assimilation/rationality) actually operate”. Therefore, we can further pay attention to how "trigger-(endogenous) activation" promotes the "diffusion" of actions, which requires a discussion of its connection with "cluster": when there are many active individuals in the network, the state ai(t)=1 of actor i can use the "trigger-(endogenous) activation" mechanism to promote the participation of active members in the neighborhood (not acting at the previous moment), thereby achieving small-scale collective actions in the neighborhood/social circle of individual i. This process is the explanation for "clusteriness" - it often starts with "small groups" and gradually spreads to the overall social network.\\

To sum up, we concretize the rational decision-making conditions/social assimilation conditions at the micro level as a “trigger-(endogenous) activation” mechanism, and based on the characteristics of social networks (limited social circles for individuals), we explain that meso/macro “cluster” phenomena can emerge under this mechanism: this can be used as a general answer to “how does action participation diffuse” in this part. Combined with the previous analysis of the four action factors, the spontaneous collective action framework can bridge the micro-macro facts and provide a more intuitive "(diffusion) mechanism explanation" for the "formation conditions".

\subsection{Synergistic Cycle Theory: Endogenous Transformation}
To explain the periodic ebb and flow of specific collective actions (eg Memorial Day marches/strikes, etc.), we can generalize the results above. In the same way as above, we need to strictly define the core variables/parameters in the formal model, including:
\begin{enumerate}[(1)]
\item  Individual willingness coefficient $f(x(t))$ can be called "exogenous willingness to act", which is determined by environmental attributes outside the system (such as economic growth/economic activity cycle), and is often cyclical.
\item I(x(t)) represents the exogenous action mapping. When its value is 1, it can be called "exogenous action activation", which represents "some individuals are not driven by the internal reasons of the system (such as friendly neighbor actions/conventional game interests), but are evoked by external attributes and emergencies (such as year-end wage arrears, etc.)"
\item Given the first moment t1 of "there is i, $a_{i}$(t) = 1", the first moment 
 ${t_{2}}$ of "any i, $a_{i}(t)$=1", the first moment $t_{3}$ of "any i, $a_{i}(t)$ = 0", and the second moment $t_{4}$ of "there is i, $a_{i}(t)$=1" (see the figure below), we respectively call ${t_{1}}$$, (t_{1}, t_{2}]$$, (t_{2}, t_{3}]$$, (t_{3}, t_{4})$ the "exogenous activation" of collective action, The stages of "endogenous diffusion", "recession" and "silence" constitute the first round of "collective action cycle" (hereinafter referred to as the cycle/the definition of the i-th cycle and so on).
\end{enumerate}
Based on the above concepts, we summarize the two core facts revealed by the above theorem/simulation: (1) In order to trigger collective action in a long period of time, there must be synergy between exogenous action activation and exogenous action intention, that is, "in the time period when exogenous action intention f(t) is high, there will be a phenomenon of exogenous action activation $I_{i}(t)$ = 1"; (2) In each cycle, the stage of individual high exogenous action intention needs to be longer, and the four formation conditions of spontaneous collective action must be met to promote the endogenous action The former means "the value of f(t) is large enough for a long time after t satisfies $I_{i}(t)$=1 (such as (t, t+k)/k is sufficiently large). In this regard, we refer to the mechanisms reflected in (1) and (2) as “exogenous synergy” and “internal and external synergy” respectively, and explain them one by one below.\\

As for the "exogenous synergy mechanism", what it shows is: under the influence of exogenous variables x(t) such as the economic cycle/public opinion cycle, the period (solution set of t) of exogenous action activation "I(t) = 1" and high exogenous willingness "f(t)\textgreater m" has overlap (intersection). To give an example, regarding the labor collective action used for calibration in this paper, if during the period of economic crisis outbreak/high incidence of wage arrears (high exogenous willingness), a vicious event of "employer running away" occurs at the same time, resulting in a small number of resistance (activation of exogenous action), which may lead to a wider range of protests and strikes (collective action). Obviously, the exogenous will/action activations here all come from the influence of the external system, so it can be called "exogenous synergy mechanism" (the overlap of two exogenous cycles).\\

The situation is different for the “external and internal synergy mechanism”, which involves the phenomenon that after exogenous synergy leads to $a_{i}(t)$ = 1 (that is, due to the overlap of high exogenous will/action activation, leading to the emergence of initial actors/“early birds”), exogenous will f(t) can still maintain a high level, so that the “activation-(endogenous) diffusion” mechanism in the theory of spontaneity can play a full role, allowing collective action to spread. Still taking the above-mentioned strike as an example, when the resistance of a few people appeared, the majority of people, under the condition of strong exogenous will, participated in the action together out of the support/imitation of workers (social assimilation) and the interest drive of salary/remuneration (rational decision-making), so as to promote a wave of collective strikes. Obviously, the resistance of the minority and the high exogenous will of the majority belong to exogenous effects, while social assimilation/rational decision-making belongs to endogenous diffusion, so it can be called "internal and external synergy".\\

Under the joint action of the two mechanisms of endogenous synergy and internal and external synergy, due to the cyclical attributes (economic cycle, etc.) of the exogenous system itself, this characteristic can be transmitted to collective actions, causing its fluctuations. In the previous paragraph, we have explained the mechanism of the two core stages of "exogenous activation" and "endogenous diffusion". As for the "recession" and "quietness" of collective action, the former is due to the "exogenous action will" entering a lower period in a cycle, which leads to the gradual withdrawal of individuals from action; From this, we can fully understand the “endogenous transformation” of the exogenous cycle. As far as strikes are concerned, the overall narrative provided by this process is (for one cycle): when the economic cycle/anger willingness is low/high, emergencies such as wage arrears/construction site accidents occur, causing a small number of people’s rights protection actions, which are coupled with others’ higher willingness to act, forming large-scale labor protests;\\

\section{Discussion and Outlook}
According to the previous discussion, we can give the significance of this paper to collective action and economic sociology: (1) For collective action, we integrate the collective action framework of Bayesian game and threshold dynamics, and introduce cycles into the model through exogenous system effects; at the same time, we further obtain theorem/simulation results and empirical conclusions about synchronization/cycle, and propose a "spontaneous cycle" theory of collective action to explain its formation and fluctuation, so as to answer questions 1 and 2. (2) For economic sociology, we not only integrate rational decision-making and social assimilation theory (two perspectives that are often separated by economics/sociology), but also incorporate exogenous economic cycles as structural elements into social movement mechanisms (integration of structure-mechanism/two-tier system), which integrates the action explanations of economics/sociology on the micro-mechanism/macro-structure respectively.\\

However, there are still many problems that need to be addressed in this theoretical framework, including: (1) further defining the exogenous cycle involved in this paper to identify the external cycle that affects the fluctuation of a specific social movement (such as the memorial day cycle corresponding to the memorial day parade); (2) integrating random disturbances and action costs in the model, and proving the corresponding conclusions; (3) introducing the perspective of political process to analyze collective actions involving government intervention-direct control. Such problems rely on more first-hand cases and field materials, and require more rigorous verification through the collection of microscopic data, which awaits follow-up attempts and research.\\

\bibliographystyle{IEEEtran}
\bibliography{ref}

\end{document}